\documentclass[11pt]{article}
\usepackage{amsmath,amsthm,amssymb,amscd}

\usepackage{hyperref}
\usepackage{url}
\usepackage{geometry}
\usepackage{tikz-cd}
\usepackage{booktabs} 
\usepackage{graphicx} 
\usepackage{multirow}
\usepackage{placeins}
\usepackage{tikz}
\usepackage{xcolor}
\usepackage{authblk}
\usetikzlibrary{calc,positioning,3d}
\usepackage[normalem]{ulem} 
\usepackage{algorithm}
\usepackage[font=footnotesize]{caption}
\usepackage{subcaption}
\usepackage{algpseudocode}%
\algrenewcommand\textproc{}
\usepackage{imakeidx}
\usepackage{authblk}

\makeindex
\usepackage[super,comma,sort&compress]  
   {natbib}
\usepackage{xcolor}

\title{PHD-MS: Multiscale Domain Identification for Spatial Transcriptomics via Persistent Homology}
\date{}

\author[1]{Perry Beamer}
\author[1,2,*]{Zixuan Cang}

\affil[1]{Department of Mathematics, North Carolina State University, Raleigh, NC, USA}
\affil[2]{Center for Research in Scientific Computation, North Carolina State University, Raleigh, NC, USA}

\affil[*]{Correspondence: zcang@ncsu.edu}

\definecolor{forest}{rgb}{0.03, 0.47, 0.19}

\graphicspath{ {./fp_figures/} }

\newcommand{\ve}[1]{\mathbf{#1}}

\ExplSyntaxOn

\cs_new_protected:Nn \__dee_mat_rows:n
 {
  \seq_set_split:Nnn \l_tmpa_seq { ; } { #1 }
  \seq_map_function:NN \l_tmpa_seq \__dee_mat_cols:n
 }

\cs_new_protected:Nn \__dee_mat_cols:n
 {
  \clist_use:nn { #1 } { & } \\
 }
\ExplSyntaxOff

\begin{document}

\maketitle

\section*{SUMMARY}

Spatial transcriptomics (ST) measures gene expression at a set of spatial locations in a tissue. Communities of nearby cells that express similar genes form \textit{spatial domains}. Specialized ST clustering algorithms have been developed to identify these spatial domains. These methods often identify spatial domains at a single morphological scale, and interactions across multiple scales are often overlooked. For example, large cellular communities often contain smaller substructures, and heterogeneous frontier regions often lie between homogeneous domains. Topological data analysis (TDA) is an emerging mathematical toolkit that studies the underlying features of data at various geometric scales. It is especially useful for analyzing complex biological datasets with multiscale characteristics. Using TDA, we develop Persistent Homology for Domains at Multiple Scales (PHD-MS) to locate tissue structures that persist across morphological scales. We apply PHD-MS to highlight multiscale spatial domains in several tissue types and ST technologies. We also compare PHD-MS domains against ground-truth domains in expert-annotated tissues, where PHD-MS outperforms traditional clustering approaches. PHD-MS is available as an open-source software package with an interactive graphical user interface for exploring the identified multiscale domains.

\section*{KEYWORDS}

Spatial transcriptomics, Multiscale domains, Topological data analysis 

\section*{INTRODUCTION}

Spatial transcriptomics (ST) technologies map gene expression to precise tissue coordinates, generating expression measurements directly linked to spatial locations within a tissue section \cite{moses2022museum}. The resolution of these measurements varies across platforms, including multi-cellular spots, single-cell resolution, and subcellular compartments. By coupling gene expression with spatial context, ST enables the study of tissue architecture, microenvironment patterns, and cell-cell interactions with unprecedented spatial details. One major analysis task in spatial transcriptomics is domain segmentation, which aims to partition the tissue into contiguous regions that are internally homogeneous in expression and consistent with morphological structure. Spatial domains provide a structural framework for characterizing tissue organization. Within these domains, domain-specific marker genes have been identified \cite{spagcn}, oncogenesis and cancer evolution have been examined \cite{seferbekova}, and a variety of analyses in cell biology and anatomy have been conducted. Consequently, numerous clustering methods have been developed specifically for spatial domain segmentation using ST data \cite{hu2024benchmarking,benchmark}. 

Current spatial domain segmentation methods mostly include three steps: a spatial neighborhood representation usually achieved by $k$-nearest neighbor graphs or distance cut-off graphs, a mechanism to integrate expression information within the spatial context such as graph neural networks, and a partitioning step to segment spatial domains by performing clustering on the features integrating expression and spatial information \cite{benchmark}. GraphST \cite{graphst}, SCAN-IT \cite{scanit}, SpaceFlow \cite{ren2022identifying}, and STAGATE \cite{dong2022deciphering} derive low-dimensional embeddings of spots from graph neural networks and then cluster them using algorithms such as the Leiden clustering. SpaGCN \cite{spagcn} further fuses histology information into the deep learning models. Statistical approaches include BASS \cite{bass} and BayeSpace \cite{BayesSpace} that use the Bayesian framework and BANKSY \cite{banksy2024} that derives interpretable augmented spatial features.

Despite strong empirical performance, these approaches often requires a user-chosen scale or resolution parameter that controls domain granularity, and in the absence of ground truth the segmentation results can be sensitive to that choice. In addition, spatial domains are often described as disjoint sets with hard boundaries in current methods.  
However, biologically, spatial domains are naturally multiscale and heterogeneous. For example, a prominent large-scale domain could contain smaller ones such as hippocampus containing Ammon's horn which further contains the CA3 subfield and so on. A single-scale approach reveals only one level of details. In another example, in the tumor microenvironment, healthy and malignant cells mingle in the tumor frontier \cite{idc_microenv}, demonstrating the heterogeneous nature of the domain. 
Revealing the mixing patterns within domains, especially near domain boundaries, as well as uncovering their hierarchical structures, remains largely unexplored.
In practice, researchers sweep parameters and visually reconcile multiple partitions. There is a lack of systematic approach to summarize the most stable structure across scales or to quantify how domains split, merge, or interact as the scale changes.

Motivated by this gap, NeST \cite{NeST2023} adopts a multiscale hotspot strategy. It identifies local enrichments of individual genes at multiple spatial resolutions and aggregates co‑localized enrichments across genes into coexpression hotspots. The hotspot detection is carried out over a range of scales, resulting in structures that vary in size and are naturally nested. This yields a hierarchical map of spatial coexpression hotspots. This framework highlights multiscale signal, but it is not formulated as domain segmentation and does not quantify how putative domains persist or interact across scales. It also does not delineate stable cores versus heterogeneous frontier regions. To address these limitations, we introduce an approach based on topological data analysis (TDA) that summarizes cross-scale cluster evolution through persistence and quantitatively characterizes mixing patterns, thereby delineating the core and frontier regions of heterogeneous tissues.
\par Topological data analysis (TDA) \cite{wasserman2018topological} is a mathematical framework for describing the shape and organization of complex data across multiple scales and dimensions. Persistent homology (PH)\cite{edelsbrunner2002topological,zomorodian2004computing}, a central method in TDA, tracks how topological structures such as connected components, loops, and higher-dimensional cavities appear and disappear with varying geometric scales, quantifying their significance by how long each feature persists across scales. In PH, the underlying structures of data across various geometric scales are organized as a \textit{filtration}, a nested sequence of simplicial complexes constructed by gradually adding connections among discrete data points in the form of edges, triangles, and other higher dimensional simplices. 
Through the filtration, PH identifies structural features in the form of different dimensional holes with connected components as 0-dimensional holes and loops as 1-dimensional holes.  
Together, PH provides a compact summary of structural features across scales and dimensions that is inherently robust to noise which is common in biological datasets.

TDA has been used for robust clustering of omics data. The Multiscale Clustering Filtration (MCF) \cite{mcf} 
considers a sequence of clusterings ordered by resolution, from fine to coarse, and uses persistent homology to represent the extent to which the clusterings are nested or hierarchical. However, the MCF is not specifically adapted for spatial data or for transcriptomic data. For single-cell transcriptomics data, HiDef \cite{hidef} applies persistent homology to study prominence of cellular communities at multiple resolutions. HiDef was developed to analyze non-spatial scRNA-seq data, and so does not consider spatial relationships between cell communities. For spatial data, a recent method TopACT \cite{Benjamin2024} employed multi-parameter PH to annotate individual cell-types using subcellular resolution spatial transcriptomic data. MCIST \cite{cotrelltranscriptomics} achieved improved clustering performance in spatial transcriptomics data by encoding the multiscale interactions between cells using persistent Laplacians.

In this paper, we introduce the concept of a multiscale domain, which relate domains identified at small and large scales. A multiscale domain consists of a homogeneous core, where spots are often assigned to the same domain across scale parameters, and heterogeneous frontier regions, where spots are less reliably included. To identify multiscale domains, we introduce a TDA pipeline, called Persistent Homology for Domains at Multiple Scales (PHD-MS). PHD-MS builds a spatial cluster filtration to identify connections between tissue domains at multiple spatial scales and compute PH to identify prominent multiscale patterns. Our multiscale approach (1) locates domains which remain stable across multiple spatial resolutions, and (2) identifies interactions between domains by considering their similarity across scales. In addition, we have implemented a point-and-click visualization utility. When only interested in visualizing the PHD-MS domains around a single point in a tissue, one can utilize this utility to plot all domains containing this point. 

We demonstrate the utility of the multiscale domain framework through several case studies. In the Visium mouse brain data, PHD-MS resolves the hierarchical organization of subregions and highlights nested structures within the tissue. When applied to a Visium breast cancer data set, PHD-MS distinguishes stable and unstable regions of the tumor microenvironment, which correspond to mixed border regions between malignant and healthy tissue. Integrated with differential gene expression analysis, PHD-MS further reveals scale-dependent transcriptional patterns within tumors.
Quantitative benchmarking across multiple Visium datasets with ground-truth annotations shows that PHD-MS consistently improves domain identification over single-scale analyses. To support this evaluation, we extend normalized mutual information (NMI) to multiscale clusters and introduce a Wasserstein distance-based metric that explicitly accounts for spatial context. Finally, we demonstrate that PHD-MS generalizes to single-cell–resolution spatial transcriptomics, capturing coherent multiscale domains in datasets from MERFISH and osmFISH.

\section*{RESULTS}

\subsection*{Overview of PHD-MS workflow}
\begin{figure}
\centering
\includegraphics[width=7in]{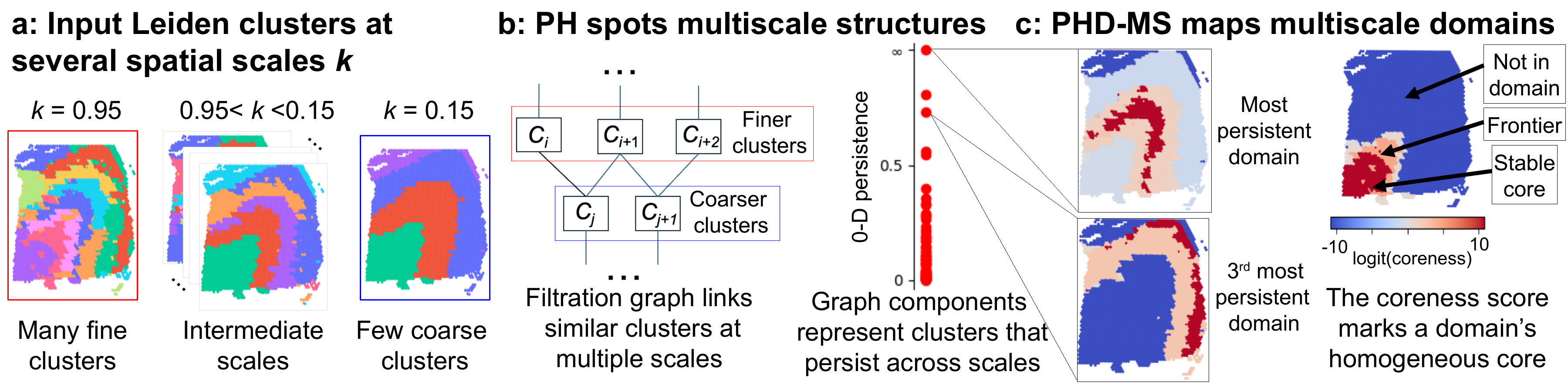}

\caption{PHD-MS method overview. \textbf{a} PHD-MS considers a sequence of domain segmentations using multiple scale parameters.
\textbf{b} A filtration function compares the overlap between clusters, creating a graph that encodes multiscale connectivity of tissue domains. Persistent homology identifies connected components, which represent prominent multiscale communities. These persistent communities are visualized using a persistence diagram.
\textbf{c} Using persistent homology, PHD-MS maps multiscale domains. Each point receives a coreness score, which represents its stability within the multiscale domain.}
\label{fig:Method-Overview}
\end{figure}
\par
PHD-MS analyzes multiscale ST clusters in three stages, aiming to identify links between domains at multiple scales. First, the data are clustered at a sequence of resolution parameters, each of which assigns every cell or spot to a spatial domain. Next, these clustering results are organized as a weighted graph whose nodes represent clusters obtained at different resolutions. This graph serves as input for the final stage, in which persistent homology (PH) is computed to identify domains that remain stable across scales. The resulting domain maps provide a multiscale visualization of tissue architecture and highlight how stable and heterogeneous regions are organized across resolutions.
\par In the first step, we consider a sequence of scale parameters $k_1,\dots,k_n$. Each scale parameter yields a clustering as a collection of disjoint domains, whose sizes typically increase as the parameter decreases. In this way, small domains grow and merge into larger domains as the scale parameter decreases (Fig. \ref{fig:Method-Overview}a).  
\par In the second step, a weighted graph is constructed to represent the connections between domains across consecutive scales. To quantify these connections, we measure the pairwise overlap between domains at each pair of neighboring scales $k_i$ and $k_{i+1}$, assigning a dissimilarity score $f$ to each pair of clusters (Fig. \ref{fig:Method-Overview}b).  Values of $f$ lie between 0 and 1, where two identical domains receive $f=0$, partially overlapping domains receives a score inversely proportional to the number of shared cells, and totally disjoint domains receive $f=1$. In this graph, each node represents a domain, and edges connect domains that share many of the same cells, weighted by their dissimilarity score. The resulting structure encodes the correspondence between spatial domains across scales.
\par In the final step, we apply PH to this weighted graph to characterize the hierarchical connections among multiscale domains. In this filtration, all nodes (domains) are present from the start, and edges are added sequentially according to their dissimilarity scores. Beginning with only edges of zero dissimilarity, additional connections are introduced in increasing order of dissimilarity. During this process, we record the values of dissimilarity score at which connected components merge. These merging events mark the scale at which previously distinct domains become unified. Components that persist over a wide range of dissimilarity values correspond to stable, recurrent domain structures that remain consistent across scales. A persistence diagram summarizes the persistence of each component (\ref{fig:Method-Overview}b).
We then project the persistent components back onto the original tissue to map multiscale domains. Each multiscale domain is represented as a set of per-spot scores, which we call a ``coreness'' score (Fig. \ref{fig:Method-Overview}c). These scores range between 0 and 1 and measure how strongly a spot belongs to the domain's stable core.  Spots with high coreness are assigned to the same underlying cluster across many scale parameters and form the domain’s homogeneous core, whereas spots with lower coreness belong to the heterogeneous outer region. Each spot receives multiple PHD-MS coreness scores, one for every multiscale domain to which it is partially assigned.

Each multiscale domain is visualized as a spatial map illustrating its stability, where higher coreness values indicate the stable core and lower values correspond to the heterogeneous outer regions (Fig.~\ref{fig:Method-Overview}c). Note that this representation is not a new clustering of the tissue into distinct domains, because multiscale domains may overlap with one another. In this way, our results can be viewed as a soft clustering of the tissue, in which domains do not rigidly assign cells to unique regions but instead represent broad, overlapping patterns of tissue organization across scales.

\subsection*{Recapitulating hierarchical organization across scales in tissues}

\begin{figure}[ht]
    \centering
    \includegraphics[width=0.89\textwidth]{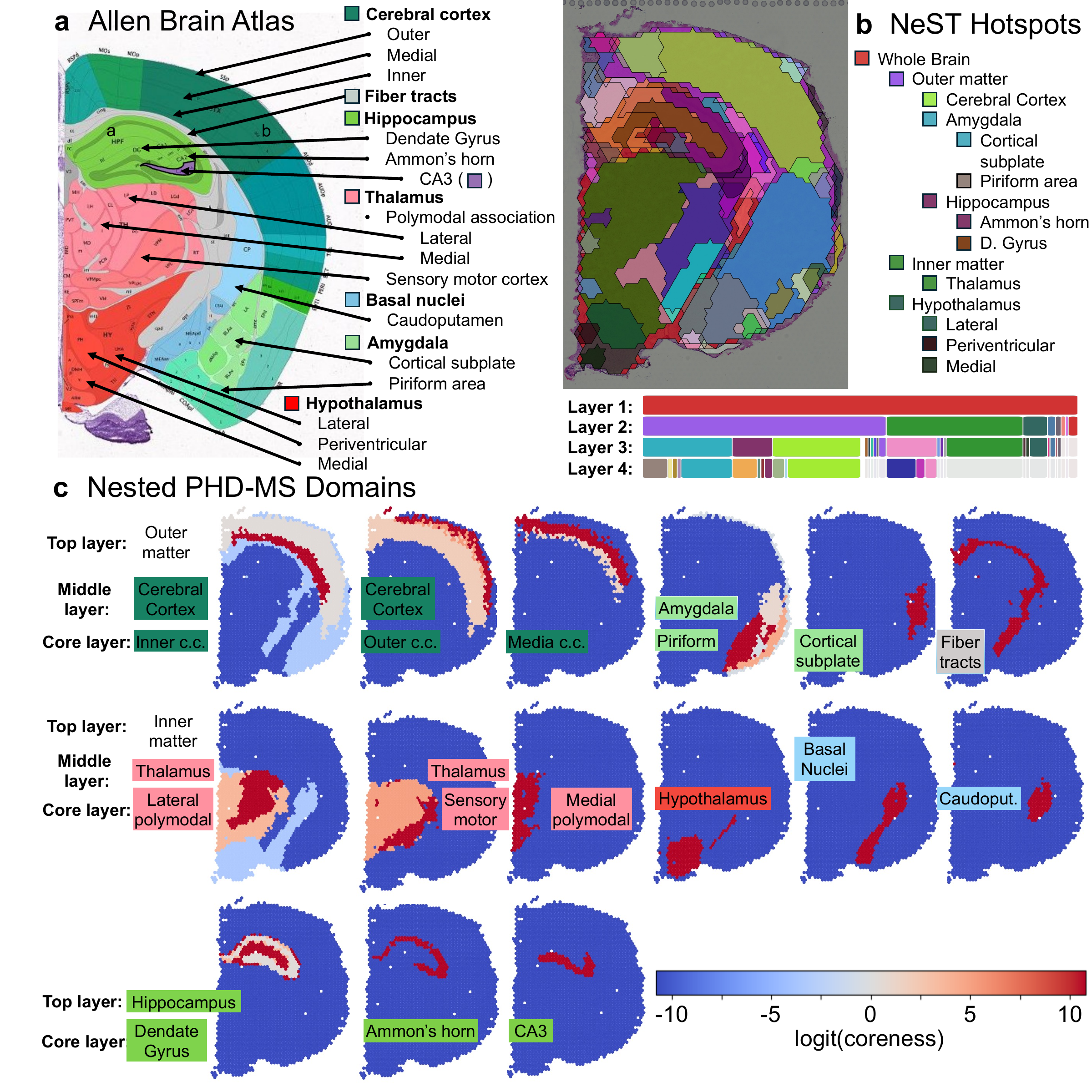}
    \caption{PHD-MS identifies stable brain structures in a Visium data of mouse brain. \textbf{a} An annotated coronal section from Allen Brain Atlas as the ground truth. \textbf{b} NeST coexpression hotspots of the Visium mouse brain data. NeST identifies 5 relevant layers of structure, colorcoded by layer (bottom), matched to the annotated regions of Allen Atlas (right).
    \textbf{c} Nested PHD-MS domains with layer labels, color coded with matched regions in 
    Allen Atlas. Top layer represents the domain at its broadest scale, including low coreness spots. Middle layer includes medium coreness spots, and Core layer forms the domain's stable core.}
    \label{fig:visium_mouse_brain}
\end{figure}

We demonstrate the utility of PHD-MS for multiscale domain representation on a Visium dataset of the coronal cross-section of mouse brain \cite{mouse_brain}. For anatomical reference, we use the Allen Brain Atlas (Fig. \ref{fig:visium_mouse_brain}a) \cite{allen,allen2}, which provides annotated coronal sections of a mouse brain. By cross-referencing the PHD-MS domains with the Allen atlas, we identify the most important fine- and coarse-scale structures in brain tissue. For comparison, NeST coexpression hotspots provide an alternative multiscale perspective (Fig. \ref{fig:visium_mouse_brain}b). Prominent PHD-MS domains recapitulate the hierarchical histological structure of the mouse brain (Fig. \ref{fig:visium_mouse_brain}c).

PHD-MS captures the organization of the coronal mouse brain across scales. At the highest level, it separates the tissue into outer cortical and inner subcortical regions. At the intermediate scale, PHD-MS recovers all major anatomical divisions annotated in the Allen Brain Atlas, including the cerebral cortex, fiber tracts, hippocampus, thalamus, basal nuclei, amygdala, and hypothalamus, and resolves specialized subregions within most of these regions (Fig. \ref{fig:visium_mouse_brain}c). In addition to regional identities, PHD‑MS distinguishes domains with diverse geometries, including compact, contiguous regions like the thalamus and narrow, elongated structures such as fiber tracts and laminar bands of the cerebral cortex. Together, these results show that PHD-MS recapitulates hierarchical tissue architecture while capturing both broad compartments and fine structures.

In comparison, NeST coexpression hotspots separate the brain into outer cortex and inner subcortical regions and, within these layers, align with major anatomical formations including the cerebral cortex, amygdala, hippocampus, thalamus, and hypothalamus. NeST further recovers several specialized subregions, especially within the hypothalamus, amygdala, and hippocampus. However, some regions do not neatly correspond to Allen Brain Atlas annotations, particularly along the interface between outer cortex and inner brain. Relative to NeST, PHD‑MS identifies several structures that NeST does not at both the intermediate and core scales. At the intermediate scale, PHD‑MS identifies every major annotated region, whereas NeST omits fiber tracts and basal nuclei. At the core scale, PHD‑MS resolves additional subregions within the cerebral cortex and thalamus. NeST uniquely identifies certain hypothalamic subregions and most effectively characterizes compact, contiguous formations, but it tends to underrepresent long, thin domains that PHD-MS captures more clearly.

\FloatBarrier

\begin{figure}
\centering
\includegraphics[width=0.9\textwidth]{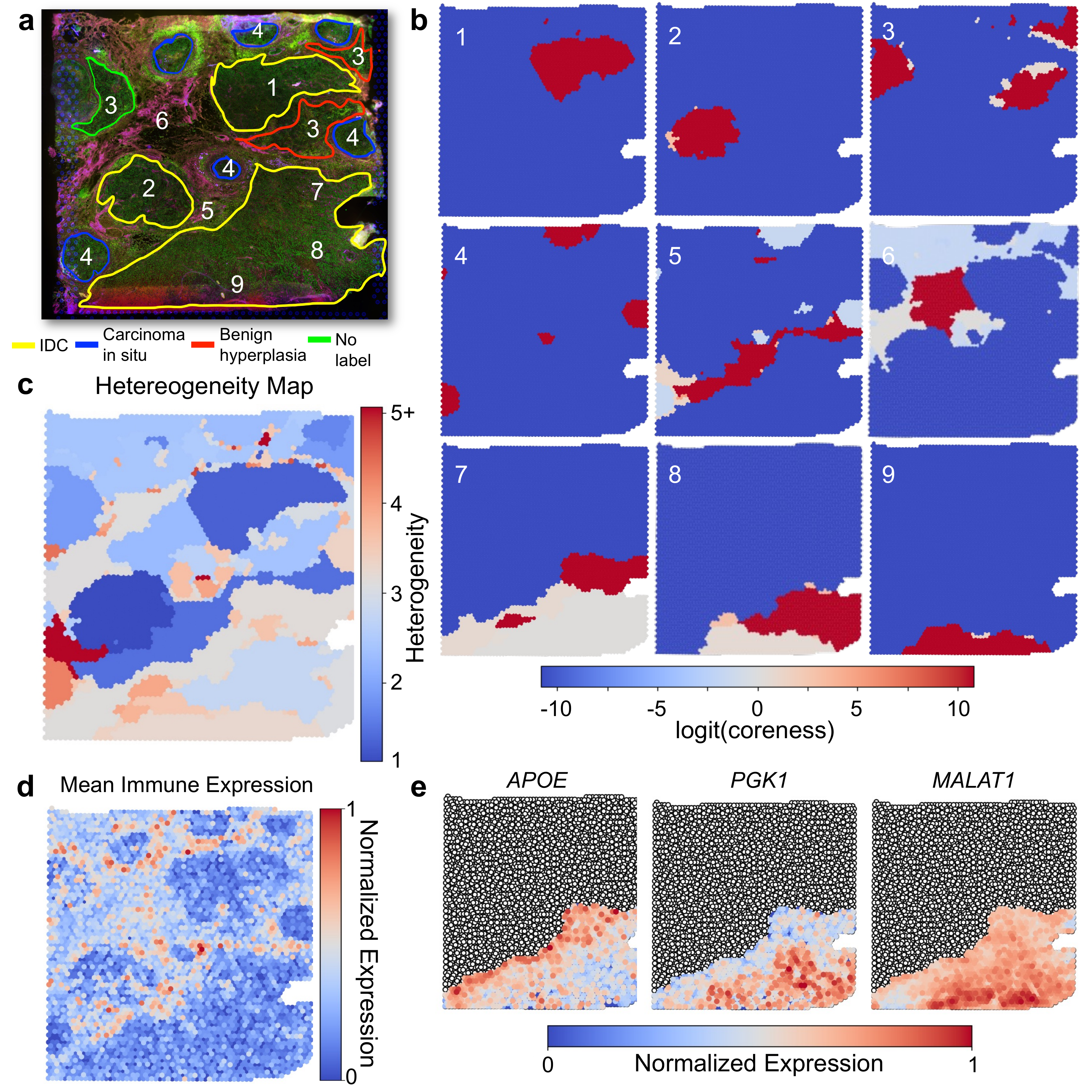}

\caption{PHD-MS analyzes tumor microenvironment morphology.
\textbf{a} Expert annotated tumors from immunofluorescent image, with anti-CD3 intensity in pink, DAPI intensity in green. IDC outlined in yellow, carcinoma \textit{in situ} outlined in blue, benign hyperplasia outlined in red, and unclassified tumor in green. 
\textbf{b} Significant domains identified by PHD-MS. Regions are numbered to correspond with the labels in \textbf{a}. \textbf{c} PHD-MS heterogeneity map highlights highly heterogeneous regions. The heterogeneity score approximates the number of persistent domains that contain each spot.
\textbf{d} Mean normalized immune expression from a set of immune genes (\textit{PTPRC}, \textit{CD4}, \textit{CD8A}, \textit{CD14}, \textit{CD68}, \textit{IGHG3}). 
\textbf{e} Normalized expression of key oncogenes \textit{APOE}, \textit{PGK1}, \textit{MALAT1} in the lower IDC. From DGE analysis, \textit{APOE} is overexpressed in region 7 relative to 8 and 9, \textit{PGK1} is overexpressed in 8, and \textit{MALAT1} is overexpressed in 9.
}\label{fig:tumor_figure}

\end{figure}
\subsection*{Multiscale domain stability reveals tumor heterogeneity}
The tumor microenvironment (TME) is highly heterogeneous, particularly along boundaries between tumors and healthy tissue. These border regions may contain a diverse mixture of cell types, including healthy tissue, immune cells, invading malignant cells, and tumor-associated stromal cells \cite{tme}. Furthermore, the tumor boundary, often a thin band of cells surrounding a much larger cancerous mass, exists at a much finer spatial scale than the tumor body. 
Here, we apply PHD-MS to study the TME of an invasive ductal carcinoma (IDC), the most common form of invasive breast cancer. Like other TMEs, the IDC microenvironment is highly heterogeneous along tumor boundaries \cite{idc_microenv}, where immune cells and tumor-associated stromal cells mix with healthy tissue. This ST dataset was analyzed and annotated by the authors of BayesSpace \cite{BayesSpace}, who discovered distinct regions of tumor transcriptomic heterogeneity. We extend this analysis across multiple scales, revealing within-region heterogeneity that is not captured by single-scale clusterings.

In Fig. \ref{fig:tumor_figure}, we highlight nine PHD-MS multiscale domains that represent the main structural characteristics of the IDC tissue. Fig. \ref{fig:tumor_figure}a shows annotation of the original immunofluorescence image, where PHD-MS identified regions are labeled with a white digit, corresponding to the PHD-MS domains in Fig. \ref{fig:tumor_figure}b. For reference, the chosen PHD-MS regions achieve an NMI of 0.53 and a mean Wasserstein distance of 535.425 $\mu\text{m}$ relative to the matched ground-truth domains, while the best GraphST clustering achieves an NMI of 0.64 and a mean Wasserstein distance of 382.345 $\mu\text{m}$. While the single-scale results perform better in quantitative benchmarking, PHD-MS uniquely reveals heterogeneous structures within the tumor microenvironment that are not captured by methods optimized for annotation agreement (Fig. \ref{fig:tumor_figure}b). When PHD-MS domains are selected at the annotation scale, the method achieves an NMI of 0.604 and a mean Wasserstein distance of 259.406~$\mu\text{m}$. We discuss quantitative metrics in the next section, where we focus on quantitatively demonstrating the accuracy of PHD-MS.

In the top row of Fig. \ref{fig:tumor_figure}b, we identify three homogeneous tumor regions corresponding to the tumors in the upper part of the tissue. Each of these regions exhibits uniformly high coreness with small low-coreness frontiers, suggesting local mixing at tumor edges.. Tumor regions 1 and 2 correspond to IDC regions in immunofluorescent image, while 3 corresponds to regions of benign hyperplasia, suggesting these tumor formations are largely homogeneous in composition. Region 4 corresponds to the major regions of carcinoma \textit{in situ}. Regions 5 and 6 represent mixed healthy/frontier regions, where high-coreness cores mark predominantly healthy tissue surrounded by less-stable border regions adjacent to tumors. Notably, region 5 contains the tumors depicted in region 4 as unstable regions. Similarly, region 9 contains tumor regions 1-3 as border domains. Such interface regions are expected to enrich for immune activity or cancer-associated stromal cells. In panels 7-9, we decompose the large lower IDC tumor into three zones with distinct cores within the larger tumor, delineating local transcriptomic variation within the same broader tumor.

To elaborate on PHD-MS's morphological characterization of the IDC tissue, we conduct a differential gene expression (DGE) analysis using the regions identified in panels 1-9 of Fig. \ref{fig:tumor_figure}. After identifying highly differentially-expressed genes for each domain, we apply gene ontology enrichment (GO) analysis \cite{GO1,GO2,GO3} to identify biological processes associated with these highly-expressed genes. Tumor regions do not show significant enrichment for broad GO biological processes.. Instead, each tumor exhibits a unique profile of oncogenic transcription. Regions 5 and 6, which contain tumor border regions, express genes associated with immune activity. According to GO analysis, highly-expressed genes in regions 5-6 are associated with a wide array of immune processes, including: B- and T-cell signaling, activation, and proliferation, complement activation, tumor necrosis factor production, and several others. Fig. \ref{fig:tumor_figure}d shows mean immune expression across a set of known immune genes, which spatially coincides with regions 5 and 6. Consistent with the coreness map, the low-coreness border of region 6 shows higher immune activity than the stable core. Moreover, comparing the three subregions of the large IDC mas (regions 7-9) with DGE analysis, we identify 17 overexpressed genes in region 7 including major IDC oncogene \textit{APOE} \cite{Zunarelli2000-bg}, 144 overexpressed genes in region 8 including oncogene \textit{PGK1} \cite{Cui2024-ap}, and 27 overexpressed genes in region 9 including oncogene \textit{MALAT1} \cite{Arun2019-ys}. Fig. \ref{fig:tumor_figure}e shows the expression of these three major IDC oncogenes, noting that regions of highest expression largely correspond to each of the three identified tumor regions. Unlike the highly stable IDCs 1-2, these results suggest that the lower IDC region contains three distinct loci of transcriptomic variation, including differences in the expression of major oncogenes.

Combined with differential gene expression analysis, these results show that PHD-MS-generated multiscale domains decompose a complex tumor microenvironment into constituent parts, highlight relationships between distinct tumor regions, and identify areas of higher and lower heterogeneity within the sample and further within individual tumors.
By introducing the coreness and heterogeneity scores, PHD-MS quantifies and characterizes transcriptional variation within the tumor microenvironment.

\subsection*{Multiscale integration improves single-resolution clustering accuracy}

\begin{figure}
    \centering
    \includegraphics[width=0.9\textwidth]{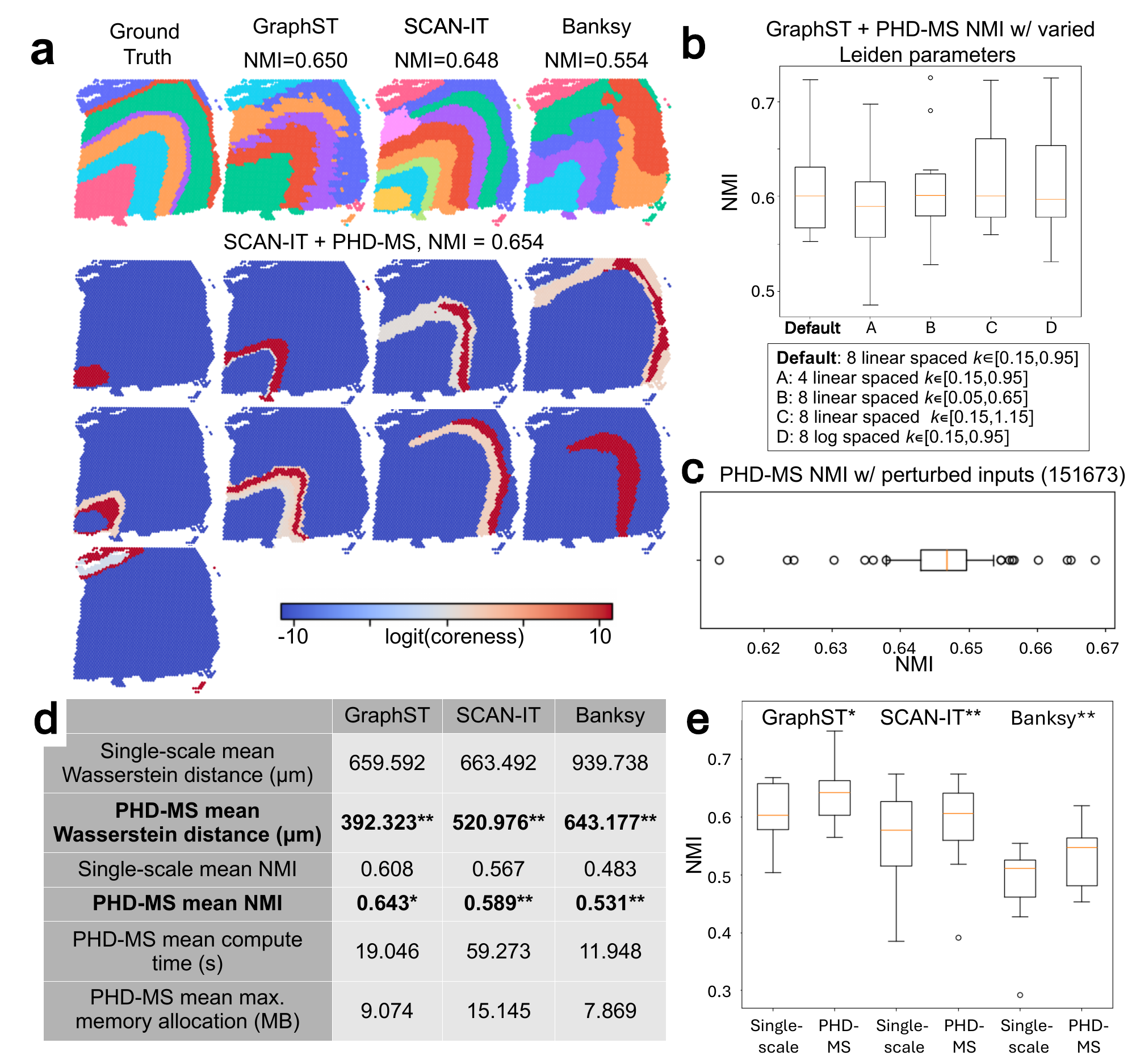}
    \caption{Quantitative benchmarking on Visium DLPFC. \textbf{a} Method comparison on slice 151673, including ground truth annotations, single-scale segmentations from GraphST, SCAN-IT, and Banksy, and a collection of PHD-MS domains generated using SCAN-IT inputs across scales.
\textbf{b} Box plots comparing NMI of PHD-MS results (constructed with GraphST embeddings) using several sets of input Leiden resolutions, where the default scheme produces the greatest median with the least negative variance.
\textbf{c} Box plot comparing PHD-MS NMI across 100  GraphST inputs generated from unique random seeds on slice 151673. Majority of results remain within 0.01 NMI across simulated perturbations. \textbf{d} Table comparing PHD-MS performance across input clustering methods. Regardless of input type, PHD-MS reports statistically significant improvement over single-scale methods with low computational cost. \textbf{e} Box plot of PHD-MS NMI across methods. PHD-MS outperforms single-scale approaches across the board. In \textbf{d} and \textbf{e}, * indicates $p<0.05$ and ** indicates $p<0.01$ statistically significant improvements over single-scale results in Wilcoxon signed-rank test.}
    \label{fig:quant-benchmark}
\end{figure}

PHD-MS domains are unique in two ways: (1) Each domain assigns a coreness score between 0 and 1 to each spot, and (2) domains may overlap. Thus, each multiscale domain is represented by a per-spot coreness score vector. In contrast, for domain segmentation at a single-scale (e.g. segmentations by GraphST, SCAN-IT, or Banksy), each transcriptomic spot is assigned to a unique domain. Such domains can be encoded as binary vectors indicating whether a spot belongs to the domain. Hereafter, we refer to these binary partitions as \textit{single-scale} or \textit{binary} domains, and to PHD-MS outputs as \textit{multiscale} domains.

Binary domains are typically compared using indices like the Jaccard index. If ground-truth annotations are available, metrics like Normalized Mutual Information (NMI) or Adjusted Mutual Information can be used to measure how accurately a set of binary domains matches the ground-truth labels. However, in their standard form, these metrics are not directly applicable to multiscale domains. To this end, we introduce two benchmarking metrics, one extending NMI to handle multiscale domains, and another capturing spatial relationships. First, we generalize NMI to operate on soft and potentially overlapping domains. The detailed definition is described in the section METHODS, and a derivation in the Supplementary Material shows that the generalized NMI reduces to standard NMI when inputs are binary. Second, we use the Wasserstein distance to quantify the spatial-aware differences between ground‑truth annotations and multiscale or single-scale domains. The Wasserstein metric complements NMI by providing a spatially interpretable distance, expressed in the units of the tissue coordinates (microns), that penalizes spatial displacement, and enables per‑domain evaluation. Equipped with these metrics, we quantitatively assess the accuracy of PHD‑MS against ground‑truth annotations and compare its performance against existing single‑scale methods.

We assess the performance of PHD-MS on several Visium datasets from the LIBD human dorsolateral prefrontal cortex (DLFPC) \cite{dlpfc} with ground-truth from expert annotations. For PHD‑MS, we evaluate the top subset of domains for each slice and compute NMI and Wasserstein distance relative to the annotations. We assess these metrics against three leading single-scale segmentation methods, GraphST, SCAN-IT, and Banksy. Fig. \ref{fig:quant-benchmark}a shows the ground truth segmentation alongside segmentations from each single-scale method and the top PHD-MS domains. As illustrated in Fig. \ref{fig:quant-benchmark}a, PHD-MS paired with SCAN-IT embeddings improves upon single-resolution results and more uniquely delineates the core regions of cortical layers.

We also conduct numerical experiments to demonstrate the robustness of PHD-MS to different types of input. Here, we obtain clusterings at different spatial scales by altering the resolution parameter in Leiden clustering algorithm. Since PHD-MS results are derived by scanning clusterings generated by several scale parameters, performance depends on the choice of parameter set and the quality of the upstream embeddings. In Fig. \ref{fig:quant-benchmark}b, we test several parameter schemes on the human DLPFC datasets (using GraphST embeddings), and compare their NMI scores. The default parameter schedule achieves the highest median NMI and the least negative variance. Moreover, the results remain relatively stable across alternative schedules, demonstrating robustness to parameter choices.
Fig. \ref{fig:quant-benchmark}c shows the NMI performance of PHD-MS across a series of perturbed input clusterings. We simulate perturbations by generating 100 unique GraphST embeddings of DLPFC slice 151673 from distinct random seeds. Each unique embedding produces a set of unique Leiden clusterings, which in turn produces a unique set of PHD-MS domains. Across these 100 simulations, the majority of results remain within 0.01 of the median NMI, indicating robustness to embedding variability. 

Lastly, we demonstrate that PHD-MS improves upon single-scale approaches overall (Fig. \ref{fig:quant-benchmark}d,e). For each method, we select the single-scale clustering that best matches the number of domains in ground truth, and compute its Wasserstein distance and NMI against the ground-truth annotations. We compare these against the top PHD-MS domains. On average, PHD-MS improves upon all three single-scale methods (GraphST, SCAN-IT, and Banksy), in both metrics.

Importantly, PHD-MS results increase in accuracy as the input quality increases. GraphST. on average, produces the most accurate single-scale results, and consequently PHD-MS is also most accurate using GraphST embeddings as inputs. In practice, using the strongest available embeddings yields the best multiscale outcomes. Additionally, runtime and memory profiling show that PHD‑MS is computationally efficient without excessive overhead (Fig. \ref{fig:quant-benchmark}d).

\subsection*{Identifying morphological domains at single-cell resolution}
\begin{figure}[ht]

\centering
\includegraphics[width=0.8\textwidth]{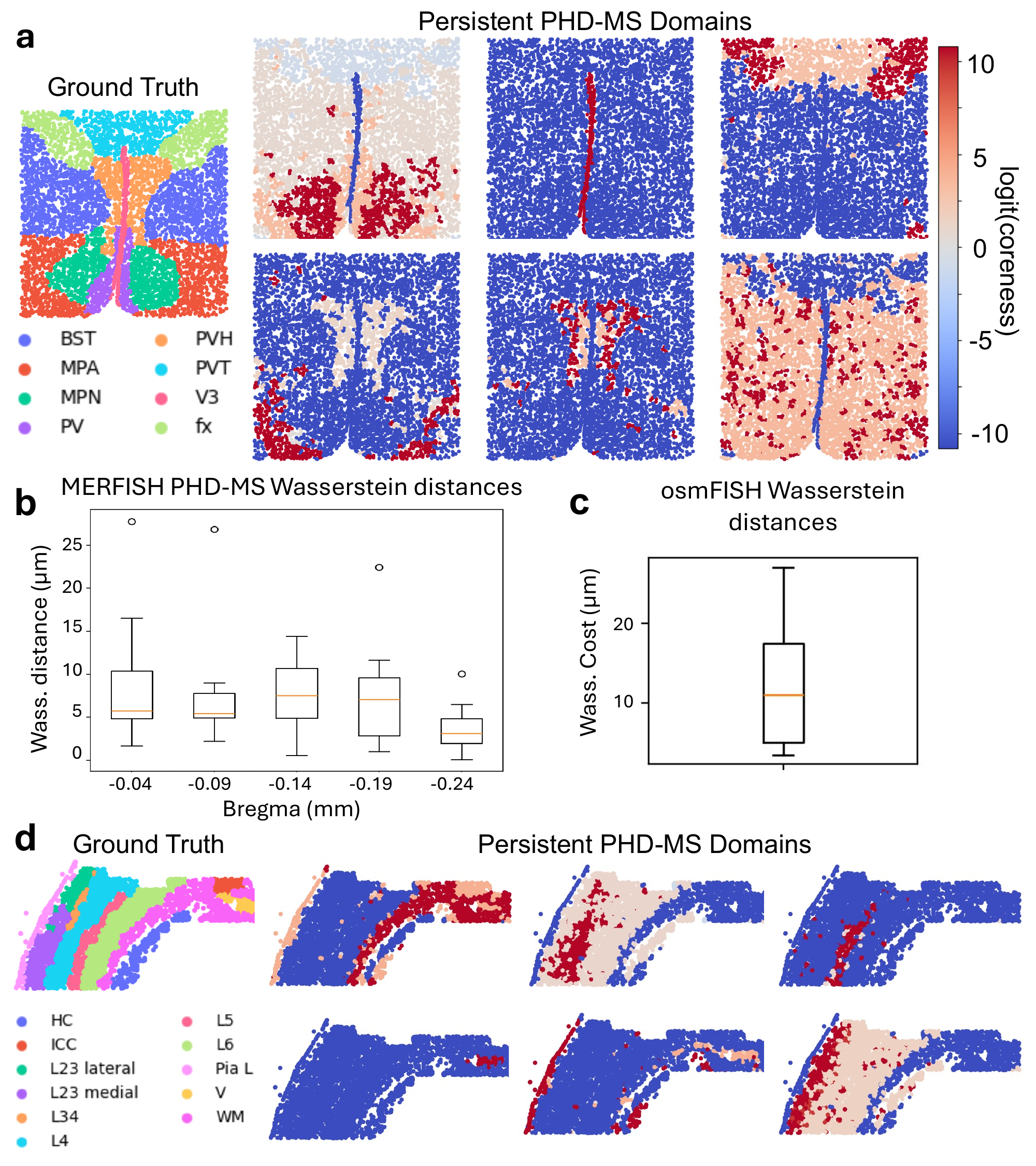}
\caption{PHD-MS reveals patterns at single-cell scale.
\textbf{a} Ground truth and persistent PHD-MS domains of the MERFISH mouse brain, bregma = -24.
\textbf{b} Boxplot of Wasserstein distance between MERFISH slices and their best-matching PHD-MS multiscale domain
\textbf{c} Boxplot of Wasserstein distance between osmFISH ground-truth domains and their best-matching PHD-MS multiscale domain.
\textbf{d} Ground truth and persistent PHD-MS domains for osmFISH}\label{fig:merfish}

\end{figure}

Several ST technologies, such as MERFISH \cite{merfish} measures a tissue's transcriptomic profile at single-cell resolution though with typical tradeoffs including limited number of assayed genes or smaller tissue areas. Here, we demonstrate that PHD-MS capably analyzes data at fine spatial-resolution. In particular, we demonstrate PHD-MS's capabilities on two representative datasets, a MERFISH mouse hypothalamic preoptic region dataset \cite{Moffit} and an osmFISH mouse somatosensory cortex \cite{osmfish} dataset.

The MERFISH mouse hypothalamic preoptic region dataset consists of several slices, labeled by distance from bregma. Each slice is annotated with ground-truth tissue domains, taken from a previous study by BayesSpace\cite{BayesSpace}. Fig. \ref{fig:merfish}a depicts the bregma -0.24 mm slice, with several persistent PHD-MS domains. These persistent PHD-MS isolate the most prominent features of the preoptic region, clearly distinguishing the midline ventricle (V3) from the rest of the hypothalamus. At top right, the thalamus is isolated, whose core aligns with the fornix (fx), and beneath it, hypothalamic tissue is identified. At bottom left, PHD-MS locates the medial preoptic area (MPA), and indicates that this region is closely associated with the periventricular hypothalamus (PVH). 
As with the Visium data, we compute the Wasserstein distance between the ground-truth domains and PHD-MS multiscale domains. Across slices, most ground-truth domains are identified with high accuracy (Fig. \ref{fig:merfish}b). The majority of ground-truth domains match an identified multiscale domain within 5 to 10 microns in Wasserstein distance, while the average distance between neighboring MERFISH spots is about 0.5 microns \cite{merfish}, indicating a deviation of approximately 10-20 spot widths. When normalized by spot spacing, this performance is comparable to that observed on Visium data in the previous section.

We also compute Wasserstein distances for the osmFISH somatosensory cortex results (Fig. \ref{fig:merfish}c). We consistently observe low Wasserstein distances, comparable to those from the MERFISH and Visium analyses. We also compare the ground-truth annotation with several PHD-MS domains (Fig. \ref{fig:merfish}d). At top left, PHD-MS identifies white matter (WM) as the core of a region that also includes all other major white matter structures: the lateral ventricle (V), corpus callosum (ICC), and Pia layer (Pia L). PHD-MS further isolates the entire somatosensory cortex, at top center panel, and subdivides it into several layers visible in subsequent domains. These results demonstrate that PHD-MS effectively captures hierarchical and spatially coherent tissue organization at single-cell resolution.

\FloatBarrier

\section*{DISCUSSION}

We introduce PHD-MS, a tool for analyzing tissue domains across spatial scales. PHD-MS is based on persistent homology, constructing a weighted cross‑scale overlap graph that ranks domain overlaps and adds edges in order of dissimilarity to merge connected components. This reveals regions that remain stable across clustering results. We provide visualizations that decompose tissue into prominent multiscale structures and identify within‑domain cores, yielding a more detailed picture of morphology than any single fixed‑scale clustering. We introduce benchmarking metrics (a multiscale NMI and a spatial Wasserstein distance) to compare against ground‑truth annotations. Finally, the tool is available as an open-source package with a user‑friendly graphical interface for exploring multiscale tissue domains.

Because of the design of the method, low-coreness regions could also arise from technical noise rather than biological heterogeneity. These cases are challenging to distinguish without biological prior knowledge, but they can be partially assessed through quality-control (QC) metrics. Specifically, we compare coreness scores with standard QC metrics such as percentage of mitochondrial genes and total counts, and inspect their spatial maps to evaluate whether low-coreness areas coincide with regions of low data quality. In the IDC example where low-coreness regions were used to interpret tumor borders, the QC maps show no localized low-quality regions, supporting the biological interpretation of low-coreness areas (SFig. 1). Furthermore, PHD-MS outputs are stable under simulated perturbations (Fig. \ref{fig:quant-benchmark}c), indicating its robustness to mild technical variation. In future analyses, especially when biological knowledge is limited, coreness scores can be compared with QC metrics, and their spatial co-localization can be used to assess biological relevance.

PHD-MS advances multiscale domain analyses beyond recent frameworks such as NeST \cite{NeST2023} and SCALE \cite{Yousefi2025.05.21.653987}, introducing several methodological innovations. As shown in Fig.~\ref{fig:visium_mouse_brain}, PHD-MS identifies hierarchical tissue structures that are not recovered by NeST. Conceptually, NeST characterizes hierarchical organization by detecting co-expression hotspots, whereas PHD-MS integrates information across resolutions to capture both hierarchical and non-hierarchical relationships among domains. In contrast to SCALE, which searches for a single optimal scale for clustering, PHD-MS aggregates information from all scales to produce a coherent set of multiscale domains. Moreover, PHD-MS introduces a quantitative coreness score that measures the membership strength of each spot within a domain, enabling numerical comparison across domains and datasets. The IDC analysis demonstrates how this score facilitates a detailed examination of intra-domain heterogeneity.

At a broader level, the need for multiscale domain identification arises because a series of single‑scale clusterings, while informative at each resolution, does not encode relationships across scales. Tissue regions interact across scales and are frequently nested. In the brain, for example, the CA3 subfield sits within Ammon’s horn, which sits within the hippocampus, which in turn belongs to the limbic system. Separate single‑scale analyses may recover each region but miss their hierarchical relations. By contrast, PHD‑MS explicitly represents domain–subdomain linkage across resolutions and visualizes these cross‑scale dependencies. Beyond hierarchy, many tissues exhibit both homogeneous regions with stable cell type composition and heterogeneous regions or region borders. In cancer microenvironments, core tumor regions dominated by malignant cells and healthy tissue are separated by frontier regions with mixed healthy and unhealthy cells, precluding strict boundaries. Traditional single‑scale clustering imposes rigid partitions and underrepresents such interfaces, whereas PHD‑MS recognizes overlapping domains across scales and marks the mixed healthy-malignant interface as an unstable subregion of both domains.

In summary, these considerations motivate the adoption of multiscale domains identification. PHD-MS complements existing spatial clustering methods by converting a collection of single-scale clustering results into an organized and interpretable map of multiscale domains. Future work could incorporate statistical investigations, such as spatial bootstraps, to quantify the uncertainty in persistences and coreness scores, and to better distinguish biological heterogeneity. Further development could enable applications to volumetric data, explore subcellular patterns in high-resolution ST data, and incorporate additional modalities such as proteomics. Finally, enhancing computational efficiency and scalability will enable applications of multiscale domain analysis to large-scale spatial transcriptomics datasets.

\section*{METHODS}

\subsection*{Cluster Filtration}
The input of PHD-MS is a collection of spatial domain clustering results of varying resolution from any clustering method of spatial transcriptomics data. Here we demonstrate the method using the clusters generated by GraphST and SCAN-IT, two highly rated methods in a recent comprehensive benchmark\cite{benchmark}. We first construct spatially-aware embeddings via GraphST or SCAN-IT. We then perform spatial domain clustering with these embeddings at a series of scale parameters using the Leiden algorithm\cite{traag2019louvain}. For the presented examples, a sequence of 8 clustering results with increasing spatial scale (larger and fewer clusters) are generated using a uniform grid of decreasing resolution parameters between $r = 0.95$ and $r=.15$ in the Leiden algorithm. The scale parameters can be easily modified by users to suit different applications.
A graph of cluster results is constructed, where each node represents a cluster at a scale $ k \in \{k_i\}_{i=1}^n$. We denote cluster $j$ (where $1\leq j \leq m_i$) at scale $k_i$ by $C_{i,j}$. Edges are drawn between all nodes from sequential scale parameters. In other words, for $i \in \{1,\dots,n-1\}$, we include an edge between each pair of clusters $(C_{i,*}\hspace{4pt};\hspace{2pt}C_{i+1,*})$ at scale $k_i$ and $k_{i+1}$.

To define a filtration on the vertex set of multi-resolution clusters, we construct a real-valued function $f$ on pairs of clusters (or edges). This function defines the order in which edges are included in persistent homology analysis. (Full definitions of persistent homology concepts are provided in the Supplementary Material). Two related filtration functions are considered, derived from the \textit{Jaccard index} and \textit{Containment index} \cite{hidef}, both which quantify similarity between clusters. The containment index filtration is intended for a sequence of clusterings where clusters are largely nested or contained within others, while the Jaccard index filtration applies more generally to any set of clusterings. We use the containment index filtration for all tissues in the results, except for the IDC in Fig. \ref{fig:tumor_figure}. We use the Jaccard index-based filtration in this case, because tumors are not necessarily organized into specialized subregions nested within larger anatomical features.
The containment index is defined by
\begin{equation}
f_C([C_{i,j},C_{i+1,l}]) = 1- \frac{|C_{i,j} \cap C_{i+1,l}|}{|C_{i,j} |}.
\end{equation}
Intuitively, $f_C$ represents the inverse proportion of nodes in $C_{i,j}$ contained in $C_{i+1,l}$.
The Jaccard index, meanwhile, is defined by:
\begin{equation}
f_J([C_{i,j},C_{i+1,l}]) = 1- \frac{|C_{i,j} \cap C_{i+1,l}|}{|C_{i,j} \cup C_{i+1,l}|}.
\end{equation}
Here, $f_J$ is the inverse fraction of the overlap between clusters over their union. In general, when$f_C$ or $f_J$ is $\approx 0$, the two clusters are highly similar, while when these functions are $\approx 1$,  two clusters are nearly entirely disjoint. 
To define a filtration, we choose the Jaccard or containment index, and extend the function to individual vertices by simply stipulating that $f_*([C_{i,j}])= 0$ for all $i,j$.

\par Intuitively, stable tissue domains across resolutions are recorded as connected components with $f$-values close to $0$. In this way, the filtration captures information about tissue domains which highly overlap as the resolution parameter varies, representing the most important underlying domains in the data.
\subsection*{Persistent Homology Results and Analysis}
For detailed definitions of concepts in persistent homology, consult the supplementary materials. We compute persistent homology using the custom-built cluster filtration via the Python package GUDHI\cite{gudhi}. We apply a union-find algorithm to recover the cells/spots which participate in each persistent connected component of the cluster filtration. These persistent connected components represent multiscale tissue domains. A coreness score is assigned to each spot in the tissue, one minus the lowest filtration value at which the spot is included in the domain. Recall that the filtration value of a cluster is determined from the Jaccard or Containment index, and a cluster is added to the connected component at this filtration value. In formal language, let $D$ denote the multiscale domain, a connected component of the filtration consisting of some set of clusters $\{C_{i,j}\}_{(i,j)\in I_D}$, and let $X$ denote the set of all tissue spots. For a spot $x$ in the domain $D$, its coreness score $c_D:X\rightarrow [0,1]$ is given by
\begin{equation}
c_D(x) = 1-\min_{C_{i,j},C_{k,l}\in D} \{f_*(C_{i,j},C_{k,l}):x\in C_{i,j}\cup C_{k,l}\}
\end{equation}
Because $x$ is a member of the domain, there exists at least one $C_{i,j} \in D$ such that $x \in C_{i,j}$, so this minimum is well-defined.
For a spot $x$ not in $D$, we set $c_D(x) = 0$. Intuitively, a low weight means $x$ is an unstable frontier member of the domain, and a high weight means $x$ is a stable core member of the domain. The multiscale domain thus consists of the pair $(D,c_D)$, the set of clusters belonging to the domain and the corresponding coreness function, where each individual domain $D$ is associated with a unique coreness function $c_D$. In practice, we typically normalize $c_D$ between $0$ and $1$ for each domain, so that the domain's stable core always has coreness score $1$:
\begin{equation}
\bar{c}_D(x) = \frac{c_D(x) - \min_y\{c_D(y)\}}{\max_y\{c_D(y)\}-\min_y\{c_D(y)\}}
\end{equation}
Recall that each domain is generated from a persistent component of the cluster filtration, so that every domain has a persistence lifetime. For a transcriptomic spot $x$, the normalized coreness score $\bar{c}_D(x)$ then represents the proportion of $D$'s lifetime where $D$ contains $x$.
\par We visualize each persistent domain $D$ by plotting normalized $\bar{c}_D$ as a scalar field on the spatial coordinates of the tissue. Spots with large $\bar{c}_D$ values are part of the domain's stable core, and spots with small $\bar{c}_D$ values are part of the domain's unstable frontier. An interactive point-and-click tool is provided for exploring the multiscale tissue morphology centered on a particular region. When the point-and-click option is selected, the user can click a point in a map of the tissue to manually visualize all the multiscale domains centered on this point.

\subsection*{Heterogeneity score}
We define the heterogeneity score, which estimates the number of unique domains that contain each transcriptomic spot. Let $D_1,\dots,D_n$ be the set of all PHD-MS domains. Recall that each $D_i$ corresponds to a persistent component in the cluster filtration. Then let $p_i$ denote the persistence lifetime (death minus birth) of $D_i$, (where we stipulate that if a domain lives forever, $p_i=1$). The heterogeneity score $h$ of transcriptomic spot $x$ is the weighted sum
\begin{equation}
h(x) = \sum_{i=1}^np_i\bar{c}_{D_i}(x)
\end{equation}
Intuitively, the heterogeneity score is high when $x$ is a core member of many highly persistent components, and is minimized when $x$ is a member of only one persistent component across scales. In this way, the heterogeneity score tracks whether $x$ is consistently assigned to the same underlying domain across scales, or if $x$ belongs to many different domains.

\subsection*{Wasserstein Distance}
The Wasserstein distance is used to compare the spatial relevance and similarity between spatial domains by representing both traditional binary domains and the multiscale domains as spatial distributions.
To compute the Wasserstein distance between two domains, we first convert each domain to a probability over the spatial coordinates of $X$. For a multiscale domain $D$, we normalize $\bar{c}_D$ to probability distributions:
\begin{equation}
\hat{c}_D(y) =  \frac{\bar{c}_D(y)}{\sum_{x\in X} \bar{c}_D(x)}
\end{equation}
In the output of existing spatial clustering algorithms and the ground truth annotations, spatial domains are often represented as traditional binary clusters, and cell-spots are not assigned a coreness score. Instead, cell-spots are either members of the binary cluster, or not. All ground-truth annotations are traditional binary clusters in this sense.
In these cases, we simply consider a uniform distribution on the spots in a binary cluster $D$
\begin{equation}
\hat{c}_D(y) = \begin{cases}
\frac{1}{|D|} & y \in D\\
0 & \text{elsewhere.}
\end{cases}
\end{equation}

Then for domains $D_1$ and $D_2$, the $2$-Wasserstein metric
\begin{equation}W_2(\hat{c}_{D_1},\hat{c}_{D_2})=\min\limits_{P\in\Gamma(\hat{c}_{D_1},\hat{c}_{D_2})}(\sum_{ij}\|x_i-x_j\|^2P_{ij})^{\frac{1}{2}}
\end{equation} depicts a spatial-aware distance between distributions $\hat{c}_{D_1}$ and $\hat{c}_{D_2}$ where $\Gamma(\hat{c}_{D_1},\hat{c}_{D_2})=\{P\in\mathbb{R}_+^{n\times n}: P\mathbf{1}_n=\hat{c}_{D_1}, P^\top\mathbf{1}_n=\hat{c}_{D_2}\}$. We compute the Wasserstein distance using the POT library\cite{flamary2021pot}.
\par In quantitative evaluations, we will compute the distance $W_2(\hat{c}_{D_1},\hat{c}_{D_2})$ where $D_2$ is a binary ground-truth domain. Then, $W_2$ measures how well a multiscale domain matches a ground-truth domain. In benchmarking, we also compute the Wasserstein distance between traditional single-scale clusters and ground-truth domains. For this purpose, we fix the resolution parameter to a single scale. In particular, we select the scale that best matches the ground-truth number of clusters. Statistical significant improvement is demonstrated by one-sided Wilcoxon signed-rank test computed using Scipy in Python, with significance thresholds $p<0.05$ and $p<0.01$.

\subsection*{Normalized Mutual Information}
Normalized mutual information (NMI) measures the shared information between two clusterings, and the NMI of a clustering and a set of ground truth labels measures the accuracy of that clustering. The standard NMI applies to binary clusters, so we have generalized it to the multiscale context. For binary clusters, our generalized version is equivalent to the standard NMI. A full derivation is provided in supplements.
\par 
Suppose we have $x_1,\dots,x_N$ transcriptomic spots. Given a subset of multiscale domains, $\{\ve D_1,\dots,\ve D_m\}$, we construct a matrix $\ve D \in [0,1]^{N\times m}$ of coreness scores, whose $ij$-th entry is the coreness score of the $i$-th transcriptomic spot in the $j$-th domain. We also normalize $\ve D$ so that its rows sum to 1:
\begin{equation}
D_{ij} = \frac{c_{D_j}(x_i)}{{\sum_{k=1}^m c_{D_{k}}(x_i)}}
\end{equation}
If $\{\ve D_1,\dots,\ve D_m\}$ are binary clusters, each spot $x_i$ belongs to one and only one $\ve D_j$. Therefore, rows of $\ve D$ will be identity vectors specifying which unique domain contains each $x_i$. 
\par 
The mutual information between two sets of multiscale domains $\{\ve D_1,\dots,\ve D_m\}$ and $\{\ve E_1,\dots,\ve E_n\}$ is given by 
\begin{equation}
\mathrm{MI}(\ve D,\ve E) = \sum_{i=1}^m\sum_{j=1}^n \frac{\ve D^\top_i\ve E_j}{N}\ln\left[\frac{N\ve D^\top_i\ve E_j}{\|\ve D_i\|_1\|\ve E_j\|_1}\right]
\end{equation}
where $\ve D_i$ denotes the $i$-th column of $\ve D$, and $\|*\|_1$ the standard $\ell_1$-norm.
\par Then the NMI is the mutual information normalized by the arithmetic mean of the Shannon's entropy $H$ of the columns of $\ve D,\ve E$, where
\begin{equation}
H(\ve D) = -\sum_{i=1}^m \frac{\|\ve D_i\|_1}{N}\ln\left(\frac{\|\ve D_i\|_1}{N}\right)
\end{equation}
Then
\begin{equation}
\mathrm{NMI}(\ve D,\ve E) = \frac{\mathrm{MI}(\ve D,\ve E)}{1/2\left[H(\ve D)+H(\ve E)\right]} 
\end{equation}
\par In quantitative evaluations, we typically compute $\mathrm{NMI}(\ve{D},\ve{G})$ where $\ve G$ is the matrix of binary ground-truth domains. Then, the NMI measures how well a set of multiscale domain matches the ground-truth domains. For benchmarking, we construct $\ve D$ from a subset of all possible PHD-MS domains, selecting a subset that maximizes the NMI. In benchmarking, we also compute the NMI between single-scale clusters and ground-truth domains. For this purpose, we fix the resolution parameter to a single scale. In particular, we select the scale that best matches the ground-truth number of clusters. Statistical significant improvement is demonstrated by one-sided Wilcoxon signed-rank test computed using Scipy in Python, with significance thresholds $p<0.05$ and $p<0.01$.

\subsection*{Differential Gene Expression}
We conduct a differential gene expression study on the Visium IDC dataset, using the 9 PHD-MS multiscale domains in Fig. 3 as categories. We assign each spot to a unique region according to its maximum coreness score across all 9 domains. For each multiscale domain, we collect all genes differentially expressed with log-fold change $>1.5$ and $p < 0.001$ according to a standard $t$-test (the rank-genes-group test implemented in the Scanpy package). In each individual domain, the expression of each gene is tested against the expression of this gene across all other domains. These highly-expressed genes are fed to a Gene Ontology Enrichment\cite{GO1} analysis, which identifies biological processes associated with these highly-expressed genes in each domain. Each significant biological process is statistically overrepresented in the set of highly-expressed genes according to Fisher's Exact test with $p<0.05$.
\par We also conduct differential gene expression studies restricted to the set of regions annotated '7', '8', and '9'. These regions border one another, and differential gene expression identifies how much oncogenic expression differs between these continuous regions. We collect all genes differentially expressed with log-fold change $>0.75$ and $p<0.001$ according to a $t$-test. Overexpressed genes are cross-referenced with the OncoDB database to identify IDC-associated genes. 

\subsection*{Data Availability}

The Visium mouse brain data \cite{mouse_brain} and MERFISH data\cite{Moffit} was accessed via the Squidpy Python package. Visium invasive ductal carcinoma data \cite{bass} is available at \url{https://www.10xgenomics.com/datasets}. Visium human dorsaleteral prefrontal cortex data \cite{dlpfc} is available from \url{https://figshare.com/articles/dataset/Visium_DLPFC_preprocessed}. OSMFISH data \cite{osmfish} is available from \url{https://figshare.com/articles/dataset/osmFISH_datasets}.
\subsection*{Code Availability}

The open-source implementation of PHD-MS (with example usage) is available on GitHub at https://github.com/pzbeamer/phd-ms.

\section*{ACKNOWLEDGMENTS}

This work was supported by NSF grant DMS2151934 and NIH grant R01GM152494.

\section*{AUTHOR CONTRIBUTIONS}

Conceptualization, P.B. and Z.C.; methodology, P.B.; investigation, P.B. and Z.C.; writing-–original draft, P.B.; writing-–review \& editing, P.B. and Z.C.; funding acquisition, Z.C.; resources, Z.C.

\section*{DECLARATION OF INTERESTS}

The authors declare no competing interests.

\bigskip


\newpage
\begin{thebibliography}{40}
\providecommand{\natexlab}[1]{#1}
\providecommand{\url}[1]{\texttt{#1}}
\providecommand{\href}[2]{#2}
\providecommand{\path}[1]{#1}
\providecommand{\DOIprefix}{doi:}
\providecommand{\ArXivprefix}{arXiv:}
\providecommand{\URLprefix}{URL: }
\providecommand{\Pubmedprefix}{pmid:}
\providecommand{\doi}[1]{\href{http://dx.doi.org/#1}{\path{#1}}}
\providecommand{\Pubmed}[1]{\href{pmid:#1}{\path{#1}}}
\providecommand{\BIBand}{and}
\providecommand{\bibinfo}[2]{#2}
\ifx\xfnm\undefined \def\xfnm[#1]{\unskip,\space#1}\fi
\makeatletter\def\@biblabel#1{#1.}\makeatother
\bibitem[{Moses and Pachter(2022)}]{moses2022museum}
\bibinfo{author}{Moses, L.}, and \bibinfo{author}{Pachter, L.} (\bibinfo{year}{2022}). \bibinfo{title}{Museum of spatial transcriptomics}.
\newblock \bibinfo{journal}{Nature methods} \emph{\bibinfo{volume}{19}}, \bibinfo{pages}{534--546}.
\bibitem[{Hu et~al.(2021)Hu, Li, Coleman, Schroeder, Ma, Irwin, Lee, Shinohara and Li}]{spagcn}
\bibinfo{author}{Hu, J.}, \bibinfo{author}{Li, X.}, \bibinfo{author}{Coleman, K.}, \bibinfo{author}{Schroeder, A.}, \bibinfo{author}{Ma, N.}, \bibinfo{author}{Irwin, D.~J.}, \bibinfo{author}{Lee, E.~B.}, \bibinfo{author}{Shinohara, R.~T.}, and \bibinfo{author}{Li, M.} (\bibinfo{year}{2021}). \bibinfo{title}{Spagcn: Integrating gene expression, spatial location and histology to identify spatial domains and spatially variable genes by graph convolutional network}.
\newblock \bibinfo{journal}{Nature Methods} \emph{\bibinfo{volume}{18}}, \bibinfo{pages}{1342--1351}. \URLprefix \url{https://doi.org/10.1038/s41592-021-01255-8}. \DOIprefix\doi{10.1038/s41592-021-01255-8}.
\bibitem[{Seferbekova et~al.(2023)Seferbekova, Lomakin, Yates and Gerstung}]{seferbekova}
\bibinfo{author}{Seferbekova, Z.}, \bibinfo{author}{Lomakin, A.}, \bibinfo{author}{Yates, L.~R.}, and \bibinfo{author}{Gerstung, M.} (\bibinfo{year}{2023}). \bibinfo{title}{Spatial biology of cancer evolution}.
\newblock \bibinfo{journal}{Nature Reviews Genetics} \emph{\bibinfo{volume}{24}}, \bibinfo{pages}{295--313}. \URLprefix \url{https://doi.org/10.1038/s41576-022-00553-x}. \DOIprefix\doi{10.1038/s41576-022-00553-x}.
\bibitem[{Hu et~al.(2024)Hu, Xie, Li, Rao, Shen, Luo, Qin, Baek and Zhou}]{hu2024benchmarking}
\bibinfo{author}{Hu, Y.}, \bibinfo{author}{Xie, M.}, \bibinfo{author}{Li, Y.}, \bibinfo{author}{Rao, M.}, \bibinfo{author}{Shen, W.}, \bibinfo{author}{Luo, C.}, \bibinfo{author}{Qin, H.}, \bibinfo{author}{Baek, J.}, and \bibinfo{author}{Zhou, X.~M.} (\bibinfo{year}{2024}). \bibinfo{title}{Benchmarking clustering, alignment, and integration methods for spatial transcriptomics}.
\newblock \bibinfo{journal}{Genome Biology} \emph{\bibinfo{volume}{25}}, \bibinfo{pages}{212}.
\bibitem[{Yuan et~al.(2024)Yuan, Zhao, Lin, Zhao, Yao, Cui, Zhang and Zhao}]{benchmark}
\bibinfo{author}{Yuan, Z.}, \bibinfo{author}{Zhao, F.}, \bibinfo{author}{Lin, S.}, \bibinfo{author}{Zhao, Y.}, \bibinfo{author}{Yao, J.}, \bibinfo{author}{Cui, Y.}, \bibinfo{author}{Zhang, X.-Y.}, and \bibinfo{author}{Zhao, Y.} (\bibinfo{year}{2024}). \bibinfo{title}{Benchmarking spatial clustering methods with spatially resolved transcriptomics data}.
\newblock \bibinfo{journal}{Nature Methods} \emph{\bibinfo{volume}{21}}, \bibinfo{pages}{712--722}. \URLprefix \url{https://doi.org/10.1038/s41592-024-02215-8}. \DOIprefix\doi{10.1038/s41592-024-02215-8}.
\bibitem[{Long et~al.(2023)Long, Ang, Li, Chong, Sethi, Zhong, Xu, Ong, Sachaphibulkij, Chen, Zeng, Fu, Wu, Lim, Liu and Chen}]{graphst}
\bibinfo{author}{Long, Y.}, \bibinfo{author}{Ang, K.~S.}, \bibinfo{author}{Li, M.}, \bibinfo{author}{Chong, K. L.~K.}, \bibinfo{author}{Sethi, R.}, \bibinfo{author}{Zhong, C.}, \bibinfo{author}{Xu, H.}, \bibinfo{author}{Ong, Z.}, \bibinfo{author}{Sachaphibulkij, K.}, \bibinfo{author}{Chen, A.}, \bibinfo{author}{Zeng, L.}, \bibinfo{author}{Fu, H.}, \bibinfo{author}{Wu, M.}, \bibinfo{author}{Lim, L. H.~K.}, \bibinfo{author}{Liu, L.}, and \bibinfo{author}{Chen, J.} (\bibinfo{year}{2023}). \bibinfo{title}{Spatially informed clustering, integration, and deconvolution of spatial transcriptomics with graphst}.
\newblock \bibinfo{journal}{Nature Communications} \emph{\bibinfo{volume}{14}}, \bibinfo{pages}{1155}. \URLprefix \url{https://doi.org/10.1038/s41467-023-36796-3}. \DOIprefix\doi{10.1038/s41467-023-36796-3}.
\bibitem[{Cang et~al.(2021)Cang, Ning, Nie, Xu and Zhang}]{scanit}
\bibinfo{author}{Cang, Z.}, \bibinfo{author}{Ning, X.}, \bibinfo{author}{Nie, A.}, \bibinfo{author}{Xu, M.}, and \bibinfo{author}{Zhang, J.} (\bibinfo{year}{2021}). \bibinfo{title}{Scan-it: Domain segmentation of spatial transcriptomics images by graph neural network}.
\newblock \bibinfo{journal}{BMVC : proceedings of the British Machine Vision Conference. British Machine Vision Conference} \emph{\bibinfo{volume}{32}}.
\bibitem[{Ren et~al.(2022)Ren, Walker, Cang and Nie}]{ren2022identifying}
\bibinfo{author}{Ren, H.}, \bibinfo{author}{Walker, B.~L.}, \bibinfo{author}{Cang, Z.}, and \bibinfo{author}{Nie, Q.} (\bibinfo{year}{2022}). \bibinfo{title}{Identifying multicellular spatiotemporal organization of cells with spaceflow}.
\newblock \bibinfo{journal}{Nature communications} \emph{\bibinfo{volume}{13}}, \bibinfo{pages}{4076}.
\bibitem[{Dong and Zhang(2022)}]{dong2022deciphering}
\bibinfo{author}{Dong, K.}, and \bibinfo{author}{Zhang, S.} (\bibinfo{year}{2022}). \bibinfo{title}{Deciphering spatial domains from spatially resolved transcriptomics with an adaptive graph attention auto-encoder}.
\newblock \bibinfo{journal}{Nature communications} \emph{\bibinfo{volume}{13}}, \bibinfo{pages}{1739}.
\bibitem[{Li and Zhou(2022)}]{bass}
\bibinfo{author}{Li, Z.}, and \bibinfo{author}{Zhou, X.} (\bibinfo{year}{2022}). \bibinfo{title}{Bass: multi-scale and multi-sample analysis enables accurate cell type clustering and spatial domain detection in spatial transcriptomic studies}.
\newblock \bibinfo{journal}{Genome Biology} \emph{\bibinfo{volume}{23}}, \bibinfo{pages}{168}. \URLprefix \url{https://doi.org/10.1186/s13059-022-02734-7}. \DOIprefix\doi{10.1186/s13059-022-02734-7}.
\bibitem[{Zhao et~al.(2021)Zhao, Stone, Ren, Guenthoer, Smythe, Pulliam, Williams, Uytingco, Taylor, Nghiem, Bielas and Gottardo}]{BayesSpace}
\bibinfo{author}{Zhao, E.}, \bibinfo{author}{Stone, M.~R.}, \bibinfo{author}{Ren, X.}, \bibinfo{author}{Guenthoer, J.}, \bibinfo{author}{Smythe, K.~S.}, \bibinfo{author}{Pulliam, T.}, \bibinfo{author}{Williams, S.~R.}, \bibinfo{author}{Uytingco, C.~R.}, \bibinfo{author}{Taylor, S. E.~B.}, \bibinfo{author}{Nghiem, P.}, \bibinfo{author}{Bielas, J.~H.}, and \bibinfo{author}{Gottardo, R.} (\bibinfo{year}{2021}). \bibinfo{title}{Spatial transcriptomics at subspot resolution with bayesspace}.
\newblock \bibinfo{journal}{Nature Biotechnology} \emph{\bibinfo{volume}{39}}, \bibinfo{pages}{1375--1384}. \URLprefix \url{https://doi.org/10.1038/s41587-021-00935-2}. \DOIprefix\doi{10.1038/s41587-021-00935-2}.
\bibitem[{Singhal et~al.(2024)Singhal, Chou, Lee, Yue, Liu, Chock, Lin, Chang, Teo, Aow, Lee, Chen and Prabhakar}]{banksy2024}
\bibinfo{author}{Singhal, V.}, \bibinfo{author}{Chou, N.}, \bibinfo{author}{Lee, J.}, \bibinfo{author}{Yue, Y.}, \bibinfo{author}{Liu, J.}, \bibinfo{author}{Chock, W.~K.}, \bibinfo{author}{Lin, L.}, \bibinfo{author}{Chang, Y.-C.}, \bibinfo{author}{Teo, E. M.~L.}, \bibinfo{author}{Aow, J.}, \bibinfo{author}{Lee, H.~K.}, \bibinfo{author}{Chen, K.~H.}, and \bibinfo{author}{Prabhakar, S.} (\bibinfo{year}{2024}). \bibinfo{title}{Banksy unifies cell typing and tissue domain segmentation for scalable spatial omics data analysis}.
\newblock \bibinfo{journal}{Nature Genetics} \emph{\bibinfo{volume}{56}}, \bibinfo{pages}{431--441}. \URLprefix \url{https://doi.org/10.1038/s41588-024-01664-3}. \DOIprefix\doi{10.1038/s41588-024-01664-3}.
\bibitem[{Kim et~al.(2020)Kim, Chung, Kim, Woo, Ahn and Park}]{idc_microenv}
\bibinfo{author}{Kim, M.}, \bibinfo{author}{Chung, Y.~R.}, \bibinfo{author}{Kim, H.~J.}, \bibinfo{author}{Woo, J.~W.}, \bibinfo{author}{Ahn, S.}, and \bibinfo{author}{Park, S.~Y.} (\bibinfo{year}{2020}). \bibinfo{title}{Immune microenvironment in ductal carcinoma in situ: a comparison with invasive carcinoma of the breast}.
\newblock \bibinfo{journal}{Breast Cancer Research} \emph{\bibinfo{volume}{22}}, \bibinfo{pages}{32}. \URLprefix \url{https://doi.org/10.1186/s13058-020-01267-w}. \DOIprefix\doi{10.1186/s13058-020-01267-w}.
\bibitem[{Walker and Nie(2023)}]{NeST2023}
\bibinfo{author}{Walker, B.~L.}, and \bibinfo{author}{Nie, Q.} (\bibinfo{year}{2023}). \bibinfo{title}{Nest: nested hierarchical structure identification in spatial transcriptomic data}.
\newblock \bibinfo{journal}{Nature Communications} \emph{\bibinfo{volume}{14}}, \bibinfo{pages}{6554}. \URLprefix \url{https://doi.org/10.1038/s41467-023-42343-x}. \DOIprefix\doi{10.1038/s41467-023-42343-x}.
\bibitem[{Wasserman(2018)}]{wasserman2018topological}
\bibinfo{author}{Wasserman, L.} (\bibinfo{year}{2018}). \bibinfo{title}{Topological data analysis}.
\newblock \bibinfo{journal}{Annual review of statistics and its application} \emph{\bibinfo{volume}{5}}, \bibinfo{pages}{501--532}.
\bibitem[{Edelsbrunner et~al.(2002)Edelsbrunner, Letscher and Zomorodian}]{edelsbrunner2002topological}
\bibinfo{author}{Edelsbrunner}, \bibinfo{author}{Letscher}, and \bibinfo{author}{Zomorodian} (\bibinfo{year}{2002}). \bibinfo{title}{Topological persistence and simplification}.
\newblock \bibinfo{journal}{Discrete \& computational geometry} \emph{\bibinfo{volume}{28}}, \bibinfo{pages}{511--533}.
\bibitem[{Zomorodian and Carlsson(2004)}]{zomorodian2004computing}
\bibinfo{author}{Zomorodian, A.}, and \bibinfo{author}{Carlsson, G.} (\bibinfo{year}{2004}).
\newblock \bibinfo{title}{Computing persistent homology}.
\newblock In: \bibinfo{booktitle}{Proceedings of the twentieth annual symposium on Computational geometry}. ( \bibinfo{pages}{347--356}).
\bibitem[{Schindler and Barahona(2023)}]{mcf}
\bibinfo{author}{Schindler, D.~J.}, and \bibinfo{author}{Barahona, M.} (\bibinfo{year}{2023}).
\newblock \bibinfo{title}{Persistent homology of the multiscale clustering filtration}.
\newblock \href{http://arxiv.org/abs/2305.04281}{\tt arXiv:2305.04281}.
\bibitem[{Zheng et~al.(2021)Zheng, Zhang, Churas, Pratt, Bahar and Ideker}]{hidef}
\bibinfo{author}{Zheng, F.}, \bibinfo{author}{Zhang, S.}, \bibinfo{author}{Churas, C.}, \bibinfo{author}{Pratt, D.}, \bibinfo{author}{Bahar, I.}, and \bibinfo{author}{Ideker, T.} (\bibinfo{year}{2021}). \bibinfo{title}{Hidef: identifying persistent structures in multiscale `omics data}.
\newblock \bibinfo{journal}{Genome Biology} \emph{\bibinfo{volume}{22}}, \bibinfo{pages}{21}. \URLprefix \url{https://doi.org/10.1186/s13059-020-02228-4}. \DOIprefix\doi{10.1186/s13059-020-02228-4}.
\bibitem[{Benjamin et~al.(2024)Benjamin, Bhandari, Kepple, Qi, Shang, Xing, An, Zhang, Hou, Crockford, McCallion, Issa, Hester, Tillmann, Harrington and Bull}]{Benjamin2024}
\bibinfo{author}{Benjamin, K.}, \bibinfo{author}{Bhandari, A.}, \bibinfo{author}{Kepple, J.~D.}, \bibinfo{author}{Qi, R.}, \bibinfo{author}{Shang, Z.}, \bibinfo{author}{Xing, Y.}, \bibinfo{author}{An, Y.}, \bibinfo{author}{Zhang, N.}, \bibinfo{author}{Hou, Y.}, \bibinfo{author}{Crockford, T.~L.}, \bibinfo{author}{McCallion, O.}, \bibinfo{author}{Issa, F.}, \bibinfo{author}{Hester, J.}, \bibinfo{author}{Tillmann, U.}, \bibinfo{author}{Harrington, H.~A.}, and \bibinfo{author}{Bull, K.~R.} (\bibinfo{year}{2024}). \bibinfo{title}{Multiscale topology classifies cells in subcellular spatial transcriptomics}.
\newblock \bibinfo{journal}{Nature} \emph{\bibinfo{volume}{630}}, \bibinfo{pages}{943--949}. \URLprefix \url{https://doi.org/10.1038/s41586-024-07563-1}. \DOIprefix\doi{10.1038/s41586-024-07563-1}.
\bibitem[{Cotrell~Sean(????)}]{cotrelltranscriptomics}
\bibinfo{author}{Cotrell~Sean, G.-W.~W.}
\newblock \bibinfo{title}{Multiscale cell-cell interactive spatial transcriptomics analysis}.
\newblock \href{http://arxiv.org/abs/https://www.researchsquare.com/article/rs-5743704/v1}{\tt arXiv:https://www.researchsquare.com/article/rs-5743704/v1}.
\bibitem[{mou(????)}]{mouse_brain}
\newblock \bibinfo{note}{{\textit{Mouse Brain Section (Coronal)}}, Spatial Gene Expression dataset analyzed using Space Ranger 1.3.0, 10x Genomics}.
\bibitem[{all(2024)}]{allen}
\bibinfo{title}{Allen Mouse Brain Atlas [dataset]}.
\newblock \bibinfo{organization}{Allen Institute for Brain Science} (\bibinfo{year}{2024}).
\bibitem[{all(2011)}]{allen2}
\bibinfo{title}{Allen Reference Atlas -- Mouse Brain [brain atlas]}.
\newblock \bibinfo{organization}{Allen Institute for Brain Science} (\bibinfo{year}{2011}).
\bibitem[{Anderson and Simon(2020)}]{tme}
\bibinfo{author}{Anderson, N.~M.}, and \bibinfo{author}{Simon, M.~C.} (\bibinfo{year}{2020}). \bibinfo{title}{The tumor microenvironment.}
\newblock \bibinfo{journal}{Curr Biol} \emph{\bibinfo{volume}{30}}, \bibinfo{pages}{R921--R925}. \DOIprefix\doi{10.1016/j.cub.2020.06.081}.
\bibitem[{Ashburner et~al.(2000)Ashburner, Ball, Blake, Botstein, Butler, Cherry, Davis, Dolinski, Dwight, Eppig, Harris, Hill, Issel-Tarver, Kasarskis, Lewis, Matese, Richardson, Ringwald, Rubin and Sherlock}]{GO1}
\bibinfo{author}{Ashburner, M.}, \bibinfo{author}{Ball, C.~A.}, \bibinfo{author}{Blake, J.~A.}, \bibinfo{author}{Botstein, D.}, \bibinfo{author}{Butler, H.}, \bibinfo{author}{Cherry, J.~M.}, \bibinfo{author}{Davis, A.~P.}, \bibinfo{author}{Dolinski, K.}, \bibinfo{author}{Dwight, S.~S.}, \bibinfo{author}{Eppig, J.~T.}, \bibinfo{author}{Harris, M.~A.}, \bibinfo{author}{Hill, D.~P.}, \bibinfo{author}{Issel-Tarver, L.}, \bibinfo{author}{Kasarskis, A.}, \bibinfo{author}{Lewis, S.}, \bibinfo{author}{Matese, J.~C.}, \bibinfo{author}{Richardson, J.~E.}, \bibinfo{author}{Ringwald, M.}, \bibinfo{author}{Rubin, G.~M.}, and \bibinfo{author}{Sherlock, G.} (\bibinfo{year}{2000}). \bibinfo{title}{Gene ontology: tool for the unification of biology}.
\newblock \bibinfo{journal}{Nature Genetics} \emph{\bibinfo{volume}{25}}, \bibinfo{pages}{25--29}. \URLprefix \url{https://doi.org/10.1038/75556}. \DOIprefix\doi{10.1038/75556}.
\bibitem[{Consortium et~al.(2023)Consortium, Aleksander, Balhoff, Carbon, Cherry, Drabkin, Ebert, Feuermann, Gaudet, Harris, Hill, Lee, Mi, Moxon, Mungall, Muruganugan, Mushayahama, Sternberg, Thomas, Van~Auken, Ramsey, Siegele, Chisholm, Fey, Aspromonte, Nugnes, Quaglia, Tosatto, Giglio, Nadendla, Antonazzo, Attrill, dos Santos, Marygold, Strelets, Tabone, Thurmond, Zhou, Ahmed, Asanitthong, Luna~Buitrago, Erdol, Gage, Ali~Kadhum, Li, Long, Michalak, Pesala, Pritazahra, Saverimuttu, Su, Thurlow, Lovering, Logie, Oliferenko, Blake, Christie, Corbani, Dolan, Drabkin, Hill, Ni, Sitnikov, Smith, Cuzick, Seager, Cooper, Elser, Jaiswal, Gupta, Jaiswal, Naithani, Lera-Ramirez, Rutherford, Wood, De~Pons, Dwinell, Hayman, Kaldunski, Kwitek, Laulederkind, Tutaj, Vedi, Wang, D'Eustachio, Aimo, Axelsen, Bridge, Hyka-Nouspikel, Morgat, Aleksander, Cherry, Engel, Karra, Miyasato, Nash, Skrzypek, Weng, Wong, Bakker, Berardini, Reiser, Auchincloss, Axelsen, Argoud-Puy, Blatter, Boutet, Breuza, Bridge, Casals-Casas, Coudert,
  Estreicher, Livia~Famiglietti, Feuermann, Gos, Gruaz-Gumowski, Hulo, Hyka-Nouspikel, Jungo, Le~Mercier, Lieberherr, Masson, Morgat, Pedruzzi, Pourcel, Poux, Rivoire, Sundaram, Bateman, Bowler-Barnett, Bye-A-Jee, Denny, Ignatchenko, Ishtiaq, Lock, Lussi, Magrane, Martin, Orchard, Raposo, Speretta, Tyagi, Warner, Zaru, Diehl, Lee, Chan, Diamantakis, Raciti, Zarowiecki, Fisher, James-Zorn, Ponferrada, Zorn, Ramachandran, Ruzicka and Westerfield}]{GO2}
\bibinfo{author}{Consortium, T. G.~O.}, \bibinfo{author}{Aleksander, S.~A.}, \bibinfo{author}{Balhoff, J.}, \bibinfo{author}{Carbon, S.}, \bibinfo{author}{Cherry, J.~M.}, \bibinfo{author}{Drabkin, H.~J.}, \bibinfo{author}{Ebert, D.}, \bibinfo{author}{Feuermann, M.}, \bibinfo{author}{Gaudet, P.}, \bibinfo{author}{Harris, N.~L.}, \bibinfo{author}{Hill, D.~P.}, \bibinfo{author}{Lee, R.}, \bibinfo{author}{Mi, H.}, \bibinfo{author}{Moxon, S.}, \bibinfo{author}{Mungall, C.~J.}, \bibinfo{author}{Muruganugan, A.}, \bibinfo{author}{Mushayahama, T.}, \bibinfo{author}{Sternberg, P.~W.}, \bibinfo{author}{Thomas, P.~D.}, \bibinfo{author}{Van~Auken, K.}, \bibinfo{author}{Ramsey, J.}, \bibinfo{author}{Siegele, D.~A.}, \bibinfo{author}{Chisholm, R.~L.}, \bibinfo{author}{Fey, P.}, \bibinfo{author}{Aspromonte, M.~C.}, \bibinfo{author}{Nugnes, M.~V.}, \bibinfo{author}{Quaglia, F.}, \bibinfo{author}{Tosatto, S.}, \bibinfo{author}{Giglio, M.}, \bibinfo{author}{Nadendla, S.}, \bibinfo{author}{Antonazzo, G.},
  \bibinfo{author}{Attrill, H.}, \bibinfo{author}{dos Santos, G.}, \bibinfo{author}{Marygold, S.}, \bibinfo{author}{Strelets, V.}, \bibinfo{author}{Tabone, C.~J.}, \bibinfo{author}{Thurmond, J.}, \bibinfo{author}{Zhou, P.}, \bibinfo{author}{Ahmed, S.~H.}, \bibinfo{author}{Asanitthong, P.}, \bibinfo{author}{Luna~Buitrago, D.}, \bibinfo{author}{Erdol, M.~N.}, \bibinfo{author}{Gage, M.~C.}, \bibinfo{author}{Ali~Kadhum, M.}, \bibinfo{author}{Li, K. Y.~C.}, \bibinfo{author}{Long, M.}, \bibinfo{author}{Michalak, A.}, \bibinfo{author}{Pesala, A.}, \bibinfo{author}{Pritazahra, A.}, \bibinfo{author}{Saverimuttu, S. C.~C.}, \bibinfo{author}{Su, R.}, \bibinfo{author}{Thurlow, K.~E.}, \bibinfo{author}{Lovering, R.~C.}, \bibinfo{author}{Logie, C.}, \bibinfo{author}{Oliferenko, S.}, \bibinfo{author}{Blake, J.}, \bibinfo{author}{Christie, K.}, \bibinfo{author}{Corbani, L.}, \bibinfo{author}{Dolan, M.~E.}, \bibinfo{author}{Drabkin, H.~J.}, \bibinfo{author}{Hill, D.~P.}, \bibinfo{author}{Ni, L.}, \bibinfo{author}{Sitnikov,
  D.}, \bibinfo{author}{Smith, C.}, \bibinfo{author}{Cuzick, A.}, \bibinfo{author}{Seager, J.}, \bibinfo{author}{Cooper, L.}, \bibinfo{author}{Elser, J.}, \bibinfo{author}{Jaiswal, P.}, \bibinfo{author}{Gupta, P.}, \bibinfo{author}{Jaiswal, P.}, \bibinfo{author}{Naithani, S.}, \bibinfo{author}{Lera-Ramirez, M.}, \bibinfo{author}{Rutherford, K.}, \bibinfo{author}{Wood, V.}, \bibinfo{author}{De~Pons, J.~L.}, \bibinfo{author}{Dwinell, M.~R.}, \bibinfo{author}{Hayman, G.~T.}, \bibinfo{author}{Kaldunski, M.~L.}, \bibinfo{author}{Kwitek, A.~E.}, \bibinfo{author}{Laulederkind, S. J.~F.}, \bibinfo{author}{Tutaj, M.~A.}, \bibinfo{author}{Vedi, M.}, \bibinfo{author}{Wang, S.-J.}, \bibinfo{author}{D'Eustachio, P.}, \bibinfo{author}{Aimo, L.}, \bibinfo{author}{Axelsen, K.}, \bibinfo{author}{Bridge, A.}, \bibinfo{author}{Hyka-Nouspikel, N.}, \bibinfo{author}{Morgat, A.}, \bibinfo{author}{Aleksander, S.~A.}, \bibinfo{author}{Cherry, J.~M.}, \bibinfo{author}{Engel, S.~R.}, \bibinfo{author}{Karra, K.},
  \bibinfo{author}{Miyasato, S.~R.}, \bibinfo{author}{Nash, R.~S.}, \bibinfo{author}{Skrzypek, M.~S.}, \bibinfo{author}{Weng, S.}, \bibinfo{author}{Wong, E.~D.} et~al. (\bibinfo{year}{2023}). \bibinfo{title}{The gene ontology knowledgebase in 2023}.
\newblock \bibinfo{journal}{Genetics} \emph{\bibinfo{volume}{224}}, \bibinfo{pages}{iyad031}. \URLprefix \url{https://doi.org/10.1093/genetics/iyad031}. \DOIprefix\doi{10.1093/genetics/iyad031}. \href{http://arxiv.org/abs/https://academic.oup.com/genetics/article-pdf/224/1/iyad031/59147104/iyad031.pdf}.
\bibitem[{Thomas et~al.(2022)Thomas, Ebert, Muruganujan, Mushayahama, Albou and Mi}]{GO3}
\bibinfo{author}{Thomas, P.~D.}, \bibinfo{author}{Ebert, D.}, \bibinfo{author}{Muruganujan, A.}, \bibinfo{author}{Mushayahama, T.}, \bibinfo{author}{Albou, L.-P.}, and \bibinfo{author}{Mi, H.} (\bibinfo{year}{2022}). \bibinfo{title}{Panther: Making genome-scale phylogenetics accessible to all}.
\newblock \bibinfo{journal}{Protein Science} \emph{\bibinfo{volume}{31}}, \bibinfo{pages}{8--22}. \URLprefix \url{https://onlinelibrary.wiley.com/doi/abs/10.1002/pro.4218}. \DOIprefix\doi{https://doi.org/10.1002/pro.4218}. \href{http://arxiv.org/abs/https://onlinelibrary.wiley.com/doi/pdf/10.1002/pro.4218}{\tt arXiv:https://onlinelibrary.wiley.com/doi/pdf/10.1002/pro.4218}.
\bibitem[{Zunarelli et~al.(2000)Zunarelli, Nicoll, Migaldi and Trentini}]{Zunarelli2000-bg}
\bibinfo{author}{Zunarelli, E.}, \bibinfo{author}{Nicoll, J.~A.}, \bibinfo{author}{Migaldi, M.}, and \bibinfo{author}{Trentini, G.~P.} (\bibinfo{year}{2000}). \bibinfo{title}{Apolipoprotein {E} polymorphism and breast carcinoma: correlation with cell proliferation indices and clinical outcome}.
\newblock \bibinfo{journal}{Breast Cancer Res. Treat.} \emph{\bibinfo{volume}{63}}, \bibinfo{pages}{193--198}.
\bibitem[{Cui et~al.(2024)Cui, Chai, Liu and Shen}]{Cui2024-ap}
\bibinfo{author}{Cui, J.}, \bibinfo{author}{Chai, S.}, \bibinfo{author}{Liu, R.}, and \bibinfo{author}{Shen, G.} (\bibinfo{year}{2024}). \bibinfo{title}{Targeting {PGK1}: A new frontier in breast cancer therapy under hypoxic conditions}.
\newblock \bibinfo{journal}{Curr. Issues Mol. Biol.} \emph{\bibinfo{volume}{46}}, \bibinfo{pages}{12214--12229}.
\bibitem[{Arun and Spector(2019)}]{Arun2019-ys}
\bibinfo{author}{Arun, G.}, and \bibinfo{author}{Spector, D.~L.} (\bibinfo{year}{2019}). \bibinfo{title}{{MALAT1} long non-coding {RNA} and breast cancer}.
\newblock \bibinfo{journal}{RNA Biol.} \emph{\bibinfo{volume}{16}}, \bibinfo{pages}{860--863}.
\bibitem[{Maynard et~al.(2021)Maynard, Collado-Torres, Weber, Uytingco, Barry, Williams, Catallini, Tran, Besich, Tippani, Chew, Yin, Kleinman, Hyde, Rao, Hicks, Martinowich and Jaffe}]{dlpfc}
\bibinfo{author}{Maynard, K.~R.}, \bibinfo{author}{Collado-Torres, L.}, \bibinfo{author}{Weber, L.~M.}, \bibinfo{author}{Uytingco, C.}, \bibinfo{author}{Barry, B.~K.}, \bibinfo{author}{Williams, S.~R.}, \bibinfo{author}{Catallini, J.~L.}, \bibinfo{author}{Tran, M.~N.}, \bibinfo{author}{Besich, Z.}, \bibinfo{author}{Tippani, M.}, \bibinfo{author}{Chew, J.}, \bibinfo{author}{Yin, Y.}, \bibinfo{author}{Kleinman, J.~E.}, \bibinfo{author}{Hyde, T.~M.}, \bibinfo{author}{Rao, N.}, \bibinfo{author}{Hicks, S.~C.}, \bibinfo{author}{Martinowich, K.}, and \bibinfo{author}{Jaffe, A.~E.} (\bibinfo{year}{2021}). \bibinfo{title}{Transcriptome-scale spatial gene expression in the human dorsolateral prefrontal cortex}.
\newblock \bibinfo{journal}{Nature Neuroscience} \emph{\bibinfo{volume}{24}}, \bibinfo{pages}{425--436}. \URLprefix \url{https://doi.org/10.1038/s41593-020-00787-0}. \DOIprefix\doi{10.1038/s41593-020-00787-0}.
\bibitem[{Chen et~al.(2015)Chen, Boettiger, Moffitt, Wang and Zhuang}]{merfish}
\bibinfo{author}{Chen, K.~H.}, \bibinfo{author}{Boettiger, A.~N.}, \bibinfo{author}{Moffitt, J.~R.}, \bibinfo{author}{Wang, S.}, and \bibinfo{author}{Zhuang, X.} (\bibinfo{year}{2015}). \bibinfo{title}{Spatially resolved, highly multiplexed rna profiling in single cells}.
\newblock \bibinfo{journal}{Science} \emph{\bibinfo{volume}{348}}, \bibinfo{pages}{aaa6090}. \URLprefix \url{https://www.science.org/doi/abs/10.1126/science.aaa6090}. \DOIprefix\doi{10.1126/science.aaa6090}. \href{http://arxiv.org/abs/https://www.science.org/doi/pdf/10.1126/science.aaa6090}{\tt arXiv:https://www.science.org/doi/pdf/10.1126/science.aaa6090}.
\bibitem[{Moffitt et~al.(2018)Moffitt, Bambah-Mukku, Eichhorn, Vaughn, Shekhar, Perez, Rubinstein, Hao, Regev, Dulac and Zhuang}]{Moffit}
\bibinfo{author}{Moffitt, J.~R.}, \bibinfo{author}{Bambah-Mukku, D.}, \bibinfo{author}{Eichhorn, S.~W.}, \bibinfo{author}{Vaughn, E.}, \bibinfo{author}{Shekhar, K.}, \bibinfo{author}{Perez, J.~D.}, \bibinfo{author}{Rubinstein, N.~D.}, \bibinfo{author}{Hao, J.}, \bibinfo{author}{Regev, A.}, \bibinfo{author}{Dulac, C.}, and \bibinfo{author}{Zhuang, X.} (\bibinfo{year}{2018}). \bibinfo{title}{Molecular, spatial, and functional single-cell profiling of the hypothalamic preoptic region}.
\newblock \bibinfo{journal}{Science} \emph{\bibinfo{volume}{362}}, \bibinfo{pages}{eaau5324}. \URLprefix \url{https://www.science.org/doi/abs/10.1126/science.aau5324}. \DOIprefix\doi{10.1126/science.aau5324}. \href{http://arxiv.org/abs/https://www.science.org/doi/pdf/10.1126/science.aau5324}{\tt arXiv:https://www.science.org/doi/pdf/10.1126/science.aau5324}.
\bibitem[{Codeluppi et~al.(2018)Codeluppi, Borm, Zeisel, La~Manno, van Lunteren, Svensson and Linnarsson}]{osmfish}
\bibinfo{author}{Codeluppi, S.}, \bibinfo{author}{Borm, L.~E.}, \bibinfo{author}{Zeisel, A.}, \bibinfo{author}{La~Manno, G.}, \bibinfo{author}{van Lunteren, J.~A.}, \bibinfo{author}{Svensson, C.~I.}, and \bibinfo{author}{Linnarsson, S.} (\bibinfo{year}{2018}). \bibinfo{title}{Spatial organization of the somatosensory cortex revealed by osmfish}.
\newblock \bibinfo{journal}{Nature Methods} \emph{\bibinfo{volume}{15}}, \bibinfo{pages}{932--935}. \URLprefix \url{https://doi.org/10.1038/s41592-018-0175-z}. \DOIprefix\doi{10.1038/s41592-018-0175-z}.
\bibitem[{Yousefi et~al.(2025)Yousefi, Schaub, Khatri, Kaiser, Kuehl, Ly, Puelles, Huber, Prinz, Krebs, Panzer and Bonn}]{Yousefi2025.05.21.653987}
\bibinfo{author}{Yousefi, B.}, \bibinfo{author}{Schaub, D.~P.}, \bibinfo{author}{Khatri, R.}, \bibinfo{author}{Kaiser, N.}, \bibinfo{author}{Kuehl, M.}, \bibinfo{author}{Ly, C.}, \bibinfo{author}{Puelles, V.~G.}, \bibinfo{author}{Huber, T.~B.}, \bibinfo{author}{Prinz, I.}, \bibinfo{author}{Krebs, C.~F.}, \bibinfo{author}{Panzer, U.}, and \bibinfo{author}{Bonn, S.} (\bibinfo{year}{2025}). \bibinfo{title}{Scale: Unsupervised multi-scale domain identification in spatial omics data}.
\newblock \bibinfo{journal}{bioRxiv}. \URLprefix \url{https://www.biorxiv.org/content/early/2025/05/27/2025.05.21.653987}. \DOIprefix\doi{10.1101/2025.05.21.653987}. \href{http://arxiv.org/abs/https://www.biorxiv.org/content/early/2025/05/27/2025.05.21.653987.full.pdf}.
\bibitem[{Traag et~al.(2019)Traag, Waltman and Van~Eck}]{traag2019louvain}
\bibinfo{author}{Traag, V.~A.}, \bibinfo{author}{Waltman, L.}, and \bibinfo{author}{Van~Eck, N.~J.} (\bibinfo{year}{2019}). \bibinfo{title}{From louvain to leiden: guaranteeing well-connected communities}.
\newblock \bibinfo{journal}{Scientific reports} \emph{\bibinfo{volume}{9}}, \bibinfo{pages}{1--12}.
\bibitem[{Maria et~al.(2014)Maria, Boissonnat, Glisse and Yvinec}]{gudhi}
\bibinfo{author}{Maria, C.}, \bibinfo{author}{Boissonnat, J.-D.}, \bibinfo{author}{Glisse, M.}, and \bibinfo{author}{Yvinec, M.} (\bibinfo{year}{2014}).
\newblock \bibinfo{title}{The gudhi library: Simplicial complexes and persistent homology}.
\newblock In: \bibinfo{editor}{Hong, H.}, and \bibinfo{editor}{Yap, C.}, eds. \bibinfo{booktitle}{Mathematical Software -- ICMS 2014}. \bibinfo{address}{Berlin, Heidelberg}: \bibinfo{publisher}{Springer Berlin Heidelberg}.
\newblock ISBN \bibinfo{isbn}{978-3-662-44199-2} ( \bibinfo{pages}{167--174}).
\bibitem[{Flamary et~al.(2021)Flamary, Courty, Gramfort, Alaya, Boisbunon, Chambon, Chapel, Corenflos, Fatras, Fournier et~al.}]{flamary2021pot}
\bibinfo{author}{Flamary, R.}, \bibinfo{author}{Courty, N.}, \bibinfo{author}{Gramfort, A.}, \bibinfo{author}{Alaya, M.~Z.}, \bibinfo{author}{Boisbunon, A.}, \bibinfo{author}{Chambon, S.}, \bibinfo{author}{Chapel, L.}, \bibinfo{author}{Corenflos, A.}, \bibinfo{author}{Fatras, K.}, \bibinfo{author}{Fournier, N.} et~al. (\bibinfo{year}{2021}). \bibinfo{title}{Pot: Python optimal transport}.
\newblock \bibinfo{journal}{Journal of Machine Learning Research} \emph{\bibinfo{volume}{22}}, \bibinfo{pages}{1--8}.


\end{thebibliography}
\end{document}